\documentclass[prd,showpacs,preprintnumbers,amsmath,nofootinbib,amssymb,floatfix]{revtex4}
\usepackage{graphicx}
\usepackage[sort&compress]{natbib}
\usepackage{subfigure}
\usepackage{amsmath}
\usepackage{amsfonts}
\usepackage{cancel}
\usepackage{lmodern,dsfont}
\usepackage{amssymb}
\begin{document}
\arraycolsep1.5pt
\newcommand{\Ima}{\textrm{Im}}
\newcommand{\Rea}{\textrm{Re}}
\newcommand{\mev}{\textrm{ MeV}}
\newcommand{\be}{\begin{equation}}
\newcommand{\ee}{\end{equation}}
\newcommand{\ba}{\begin{eqnarray}}
\newcommand{\ea}{\end{eqnarray}}
\newcommand{\gev}{\textrm{ GeV}}
\newcommand{\nn}{{\nonumber}}
\newcommand{\dtres}{d^{\hspace{0.1mm} 3}\hspace{-0.5mm}}
\newcommand{\rts}{ \sqrt s}
\newcommand{\non}{\nonumber \\[2mm]}

\title{The $\pi \pi$ Interaction in the $\rho$ Channel in Finite Volume}

\author{Hua-Xing Chen$^{1,2}$}
\email{hxchen@ific.uv.es}
\author{E. Oset$^{1}$}
\email{oset@ific.uv.es}
\affiliation{
$^1$ Departamento de F\'{\i}sica Te\'orica and IFIC, Centro Mixto Universidad de Valencia-CSIC, Institutos de Investigaci\'on de Paterna, Aptdo. 22085, 46071 Valencia, Spain \\
$^2$ School of Physics and Nuclear Energy Engineering, Beihang University, Beijing 100191, China
}

\date{\today}

\begin{abstract}
 The aim of this paper is to investigate an efficient strategy that allows to obtain $\pi \pi$ phase shifts and $\rho$ meson properties from QCD lattice data with high precision. For this purpose we evaluate the levels of the $\pi \pi$ system in the $\rho$ channel in finite volume using chiral unitary theory. We investigate the dependence on the $\pi$ mass and compare with other approaches which use QCD lattice calculations and effective theories. We also illustrate the errors induced by using the conventional L\"uscher approach instead of a more accurate one recently developed that takes into account exactly the relativistic two meson propagators. Finally we make use of this latter approach to solve the inverse problem, getting $\pi \pi$ phase shifts from ``synthetic'' lattice data, providing an optimal strategy and showing which accuracy is needed in these data to obtain the $\rho$ properties with a desired accuracy.

\end{abstract}

\maketitle

\section{Introduction}
\label{Intro}

The determination of hadron spectra is one of the challenging tasks of Lattice QCD and many efforts are being devoted to this problem~\cite{Nakahara:1999vy,Mathur:2006bs,Basak:2007kj,Bulava:2010yg,Morningstar:2010ae,Foley:2010te,Alford:2000mm,Kunihiro:2003yj,Suganuma:2005ds,Hart:2006ps,Wada:2007cp,Prelovsek:2010gm,Lin:2008pr,Gattringer:2008vj,Engel:2010my,Mahbub:2010me,Edwards:2011jj,sasa,Prelovsek:2011im}.
As one is aiming at higher accuracy and consistency with the decay channels of the hadrons, L\"uscher's approach~\cite{luscher,Luscher:1990ux}, reconstructing phase shifts of the decay products from the discrete energy levels of the box,  is playing gradually a more important role. Other strategies formerly used, like making use of the ``avoided level crossing'', are leaving room to other methods once this criteria has been shown insufficient for resonances with a large width~\cite{Bernard:2007cm,Bernard:2008ax,misha}.

One of the hadrons that has received more attention is the $\rho$ meson. There have been a few attempts to describe the $\rho$ using L\"uscher's approach~\cite{Feng:2011ah,Aoki:2007rd,Gockeler:2008kc,Aoki:2010hn,Feng:2010es,Frison:2010ws}. On the other hand the first lattice estimate of the $\rho \to \pi \pi$ amplitude \cite{McNeile:2002fh} did not apply L\"uscher's method. The most recent work on the $\rho$ properties from the lattice point of view is the one of \cite{sasa} and we shall refer to it for comparison along the paper.  In between, in \cite{misha} L\"uscher's approach has been recently simplified and improved by keeping the full relativistic two body propagator (L\"uscher's approach keeps the imaginary part of this propagator exactly but makes approximations on the real part) and extending the method to two or more coupled channels. The method has also been applied in~\cite{mishajuelich} to obtain finite volume results from the J\"ulich model for meson baryon interaction and in \cite{alberto}, to study the interaction of the $DK$ and $\eta D_s$ system where the $D_{s^*0}(2317)$ resonance is dynamically generated from the interaction of these particles~\cite{Kolomeitsev:2003ac,Hofmann:2003je,Guo:2006fu,daniel}. The case of the $\kappa$ resonance in the $K \pi$ channel is also addressed along the lines of \cite{misha} in \cite{mishakappa} and an extension of the approach of \cite{misha} to the case of interaction of unstable particles has been made in \cite{luisroca}.

In the present work we shall use the approach of \cite{misha} in the case of the interaction of two pions in p-waves and isospin $I=1$, the $\rho$ channel.  In  the first part we shall use the chiral unitary approach to $\pi \pi$ scattering in p-waves from \cite{nsd,formfactor} and apply it to get energy levels in a finite box. We shall then see how they change with the pion mass to establish connection with lattice results which run for heavier masses than the physical pion mass. In a second step we will face the inverse problem of getting $\pi \pi$ phase shifts from the energy spectra in the box, determining the precision needed in the lattice spectra to get the phase shifts with a demanded accuracy.

 The paper proceeds as follows: Section II introduces the chiral unitary approach for $\pi \pi $ and $K \bar K $ scattering in the $\rho$ region, both in the infinite space and in finite volume. Section III applies L\"uscher formalism to get phase shifts from finite volume spectra and shows that, because the $\rho$ is basically a $q \bar q$ state, the "potential" needed in the Bethe Salpeter equations is singular, it contains a singularity in terms of a bare $\rho$ pole, but, in spite of it, L\"uscher formalism is still very efficient.
In section IV we make a study of the $m_{\pi}$ dependence of the results for the $\rho$ and we compare with the lattice data. The study shows that while some magnitudes are strongly dependent on $m_{\pi}$, the coupling $g_{\rho \pi \pi}$  is very smoothly dependent. In section V we discuss the differences between the standard L\"uscher approach and the one followed here, and we give arguments on how one can use small lattice boxes while still having accurate results using our method. Section VI is the most practical for lattice QCD practitioners. It proposes a chiral unitary approach motivated type of potential, including a CDD pole and unknown parameters which are fitted to the bulk of the lattice spectra, going beyond the L\"uscher approach of getting directly phase shifts for the particular eigenenergies of the finite box. We show that this method is very efficient, providing phase shifts in a wide range of energies, some of which would require the use of too small or too large box sizes in the direct L\"uscher method. In section VII we investigate other possible fit strategies where the data are all obtained from one or two volumes. We show that the strategy works and this can save computing time in actual QCD lattice calculations. We also show the limits to this method, since adding some new levels can make the use of the two coupled channels necessary. Finally we show some concluding remarks in section VIII.

\section{Formalism of the chiral unitary approach for the $\rho$ in infinite and finite volume.}

We shall follow here the approach of \cite{formfactor} to get the $\pi \pi$ scattering amplitude in p-waves. In this work the pion and kaon form factors were studied using the chiral Lagrangians of \cite{gasser} and those for the coupling of resonances to pairs of pseudoscalar mesons of \cite{rafael}. The work was done obtaining  a kernel, or $\pi \pi$ potential, from those Lagrangians, which contains a contact term from the lowest order Lagrangian of \cite{gasser} and another term, the explicit $\rho$ pole, with the coupling of the $\rho$ to the pions obtained from \cite{rafael}. One includes also the $K\bar K$ system and unitarizes the two channel problem using the coupled channels Bethe Salpeter equations, or the equivalent $N/D$ method. Although one can approach the problem from different perspectives, like the inverse amplitude method (IAM) \cite{Dobado:1996ps,Oller:1998hw}, or the Bethe Salpeter equations with an energy dependent potential provided by the first and second order terms of the chiral Lagrangians of \cite{gasser}, as done in \cite{borasoy}, the approach of \cite{formfactor} by separating the contact term and a bare $\rho$ pole, already accepts from the beginning the proven fact that the $\rho$ is not a $\pi \pi$ resonance, but rather a genuine state which has $\pi \pi$ as decay channel.  This has been checked thoroughly by looking at the large $N_c$ behaviour of the pole position in \cite{Pelaez:2003dy,riosjose}, where the mass of the $\rho$ stays constant as $N_c$ grows and the width decreases. This is opposite to the composite states, dynamically generated by the interaction, like the light scalar mesons, where the width grows considerably as $N_c$ increases. A different perspective is provided in \cite{Nieves:2009ez}, where it is shown that all chiral logarithms cancel out in the rho-channel, while they do not cancel for the sigma case, and they strongly influence the properties of this latter resonance. In other words, the loops from intermediate $\pi \pi$ states are very relevant for the sigma and highly unimportant for the $\rho$ case. In even simpler words we could say that the sigma is a $\pi \pi$ resonance, while the $\rho$ is a non $\pi \pi$ state which decays into this channel. This has also been proved using an extension of the Weinberg test of composite particles in \cite{aceti}. One way to take that nature into account is to allow for a CDD (Castillejo, Dalitz, Dyson) pole in the potential \cite{Castillejo:1955ed}, and this is what naturally appears by combining the Lagrangians of \cite{gasser} and \cite{rafael} and is implemented in the approach of \cite{formfactor}.

In the chiral unitary approach one uses the coupled channel Bethe-Salpeter equations in their on shell factorized form \cite{npa,ollerulf}
\begin{eqnarray}
{\bf T} &=& (1 - {\bf V} {\bf G})^{-1} {\bf V} \, .
\label{bethesal}
\end{eqnarray}
The relevant $V$-matrices for $\pi \pi$ and $K\bar K$ scattering have been studied in Ref.~\cite{formfactor}:
\begin{eqnarray}\label{eq:Vmatrix}
\nonumber V^{I=1}_{\pi \pi, \pi \pi} &=& T_{\pi^+ \pi^-, \pi^+ \pi^-} = - {2 p_\pi^2 \over 3 f^2} \Big ( 1 + {2 G_V^2 \over f^2} { s \over M_\rho^2 - s } \Big ) \, ,
\\ V^{I=1}_{\pi \pi, K \bar K} &=& \sqrt 2 T_{\pi^+ \pi^-, K^+ K^-} = - { \sqrt2 p_\pi p_K \over 3 f^2} \Big ( 1 + {2 G_V^2 \over f^2} { s \over M_\rho^2 - s } \Big ) \, ,
\\ \nonumber V^{I=1}_{K \bar K, K \bar K} &=& T_{K^+ K^-, K^+ K^-} - T_{K^+ K^-, K^0 \bar K^0} = - {p_K^2 \over 3 f^2} \Big ( 1 + {2 G_V^2 \over f^2} { s \over M_\rho^2 - s } \Big ) \, ,
\\ \nonumber V^{I=0}_{K \bar K, K \bar K} &=& T_{K^+ K^-, K^+ K^-} + T_{K^+ K^-, K^0 \bar K^0} = - {p_K^2 \over 3 f^2} \Big ( 3 + {2 G_V^2 \over f^2} { s \over M_\omega^2 - s } + {4 G_V^2 \over f^2} { s \over M_\phi^2 - s } \Big ) \, .
\end{eqnarray}
We note that these definitions have a minus sign difference with respect to \cite{formfactor} to follow the more standard notation for $V$ and $T$. We work in the isospin base, where the states are defined as:
\begin{eqnarray}
| \pi \pi \rangle _ {I=1} &=& {1\over2} | \pi^+ \pi^- - \pi^- \pi^+ \rangle \, ,
\\ | K \bar K \rangle _ {I=1} &=& {1\over\sqrt2} | K^+ K^- - K^0 \bar K^0 \rangle \, ,
\\ | K \bar K \rangle _ {I=0} &=& {1\over\sqrt2} | K^+ K^- + K^0 \bar K^0 \rangle \, .
\end{eqnarray}
Note the extra $1 / \sqrt2$ in the normalization of the pions (unitary normalization) which is adopted to account for the identity of the particles in the intermediate states. Since in this paper we study the $\rho$ meson which has isospin $I=1$, there are only two channels to be considered, $| \pi \pi \rangle _ {I=1}$ and $| K \bar K \rangle _ {I=1}$. For simplicity, we rewrite the $V$-matrix as $V_{11} = V^{I=1}_{\pi \pi, \pi \pi}$, $V_{22} = V^{I=1}_{K \bar K, K \bar K}$ and $V_{12} = V^{I=1}_{\pi \pi, K \bar K}$.

The parameters in Eq.~(\ref{eq:Vmatrix}) are also taken from Ref.~\cite{formfactor}, in particular:
\begin{eqnarray}
\nonumber G_V &=& 53 {\rm MeV} \, ,
\\ f &=& 87.4 {\rm MeV} \, .
\end{eqnarray}
But for the bare $\rho$ mass $M_\rho$ of Eqs. (\ref{eq:Vmatrix}) we choose the value $837.3$ MeV, which is slightly larger than the value $829.8$ MeV used in Ref.~\cite{formfactor}. Doing this, the phase shifts obtained are slightly better compared with the experimental data~\cite{Estabrooks:1974vu}, as shown in Fig.~\ref{fig:PhaseRead} in the next section.

The $G$-function for the two-meson propagator having masses $m$ and $M$ is defined as
\begin{eqnarray}
G (p^2) &=& i \int {d^4 q \over (2 \pi)^4} {1 \over q^2 - m^2 + i \epsilon} {1 \over (p - q)^2 - M^2 + i \epsilon} \, .
\end{eqnarray}
In Ref.~\cite{formfactor} it has been calculated in the continuum space (infinite volume) using  dimensional regularization, and the result is:
\begin{eqnarray}\label{eq:GDR}
G_i (s) &=& {1 \over 16 \pi^2} \Big ( -2 + d_i + \sigma_i(s) \log{ \sigma_i(s) + 1 \over \sigma_i(s) - 1 } \Big ) \, ,
\label{Ggasser}
\end{eqnarray}
where the subindex $i$ refers to the corresponding two-meson state, either $| \pi \pi \rangle _ {I=1}$ ($i=1$) or $| K \bar K \rangle _ {I=1}$ ($i=2$), and $\sigma_i(s) = \sqrt{1 - 4 m_i^2/s}$. The parameters $d_i$ are also taken from Ref.~\cite{formfactor} and are given by:
\begin{eqnarray}\label{dscpt}
d_1 &=& {m_K^2 \over m_K^2 - m_\pi^2} \big ( \log{m_\pi^2 \over \mu^2} + {1\over2}\log{m_K^2 \over \mu^2} + {1\over2} \big ) \, ,
\\ \nonumber d_2 &=& { - 2 m_\pi^2 \over m_K^2 - m_\pi^2} \big ( \log{m_\pi^2 \over \mu^2} + {1\over2}\log{m_K^2 \over \mu^2} + {1\over2} \big ) \, ,
\end{eqnarray}
with $\mu=M_{\rho}$. Eqs.~(\ref{dscpt}) are taken such as to guarantee a perfect matching with the results of chiral perturbation theory for the form factor of \cite{Gasser:1984ux}.

All the above formulae are for the study of the $\rho$ meson in the infinite space. To study the $\rho$ meson in the finite volume, we just need to change the $G$-function of dimensional regularization (Eq.~(\ref{eq:GDR})) by the one which is defined in the finite box of side $L$. This was deduced in  \cite{alberto} and it is given by $\tilde G(s)$, defined through:
\begin{eqnarray}\label{eq:difference}
\tilde G(s) - G(s) &=&
\lim_{q_{\rm max} \rightarrow \infty} \Big ( {1 \over L^3} \sum_{q_i}^{q_{\rm max}} I(q_i) - \int_{q<q_{\rm max}}{d^3 q \over (2\pi)^3} I(q) \Big ) \, .
\label{gtilde}
\end{eqnarray}
\begin{eqnarray}
I(q_i) =  {1 \over 2 \omega_1(\vec q) \omega_2(\vec q)} {\omega_1(\vec q) + \omega_2(\vec q) \over E^2 - (\omega_1(\vec q) + \omega_2(\vec q))^2} \, ,
\label{ifun}
\end{eqnarray}
where $\omega_{1,2}(\vec q) = \sqrt{m_{1,2}^2 + \vec q^2}$ and the discrete momenta in the sum given by $\vec q = {2 \pi \over L} \vec n ~~ ( \vec n \in \mathcal{Z}^3 )$. In this way we have just changed the integration over momenta by a sum over the discrete values of the momenta allowed by the periodic conditions in the box.

\begin{figure}[hbt]
\begin{center}
\scalebox{0.8}{\includegraphics{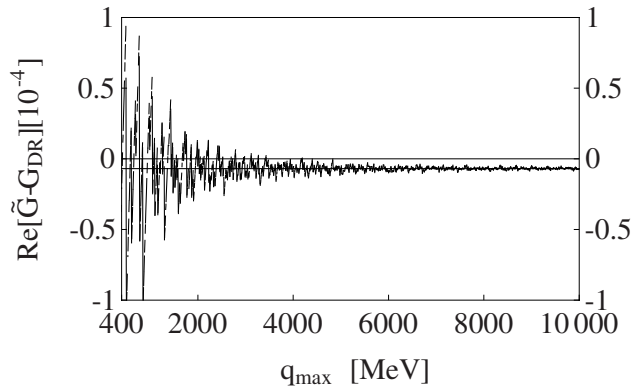}} \caption{The difference between $\tilde G(s)$ and ${\rm Re}[G_{\rm DR}(s)]$ calculated using Eq.~(\ref{eq:difference}) as a function of $q_{\rm max}$. Here we choose $L = 2.5~m_\pi^{-1}$ and E = 770 MeV.} \label{fig:difference}
\end{center}
\end{figure}

The real part of  $\tilde G(s) - G(s)$ is shown in Fig.~\ref{fig:difference} as a function of $q_{\rm max}$. We note that we have fixed $L$ to be 2.5 $m_\pi^{-1}$ and $E$ to be $770$ MeV. We find that it has a good convergence when one goes to large values of $q_{max}$. However, as already done in \cite{misha,alberto} it is more  practical and equally accurate to make an average of this quantity for smaller values of $q_{max}$ with the consequent economy in computational time.

\section{The Energy Levels in the Chiral Unitary Approach}
\label{sec:energylevels}

Using the $\tilde G$-function in Eq.~(\ref{gtilde}), we calculate the energy levels in the finite volume for different values of $L$, by looking at the poles of the $T$ matrix, which appear when the determinant of $1 - {\bf V} {\bf G}$ is zero:
\begin{eqnarray}\label{eq:EL}
{\rm Det} (1 - {\bf V} {\bf G}) = 1 - V_{11} \tilde G_1(s) - V_{22} \tilde G_2(s) + (V_{11} V_{22} - V_{12}^2) \tilde G_1(s) \tilde G_2(s) = 0 \, .
\end{eqnarray}
As we have said before, our procedure follows closely the method used in  Refs.~\cite{misha,alberto,mishakappa,Xie:2012pi
}, so here we go directly to the discussion of the numerical results. The energy levels for $\pi \pi$ $P$-wave scattering are functions of the cubic box size $L$, as well as the pion mass $m_\pi$. We shall study the dependence on these variables. The volume dependence is shown here and the pion mass dependence in the next section.

\begin{figure}[hbt]
\begin{center}
\scalebox{0.8}{\includegraphics{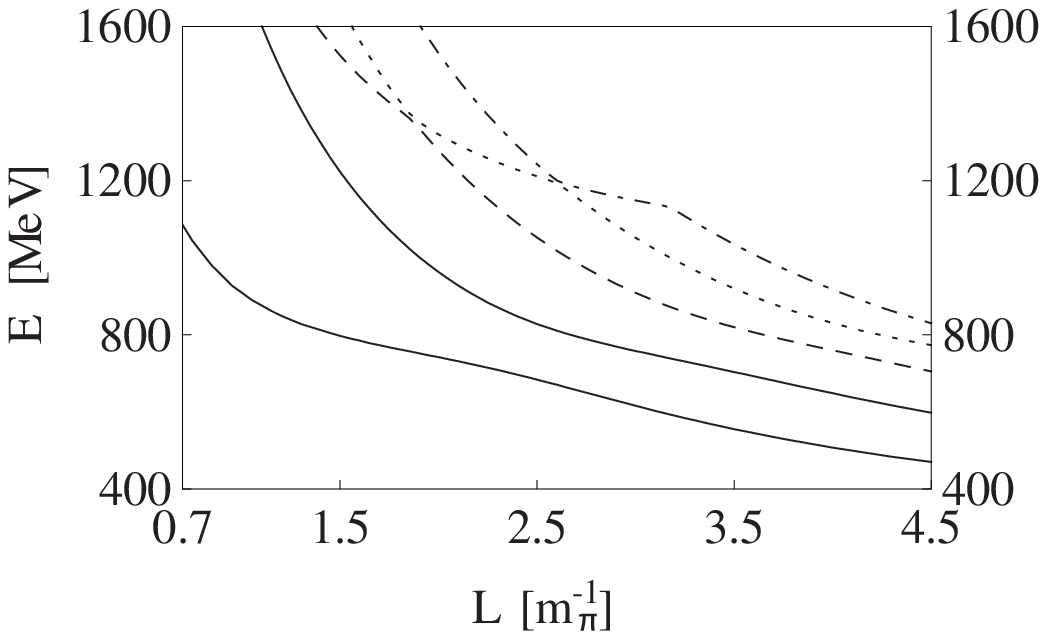}}
\scalebox{0.8}{\includegraphics{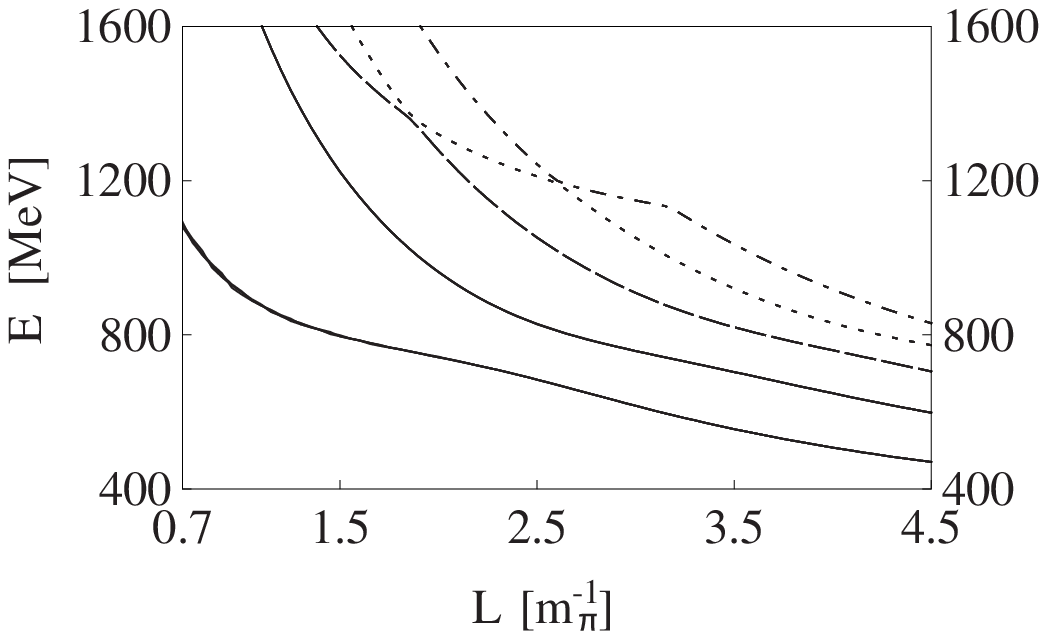}}
\caption{Energy levels as functions of the cubic box size $L$, derived from the coupled channels unitary approach of Ref.~\cite{formfactor} and using $\tilde G$ from Eq.~(\ref{gtilde}). We perform an average for different $q_{\rm max}$ values between 1200 MeV and 2000 MeV in the left figure, while show them separately in the right figure for different $q_{\rm max} = 1300$, $1500$, $1700$ and $1900$ MeV, which are almost the same.}
\label{fig:ELMix}
\end{center}
\end{figure}

\begin{figure}[hbt]
\begin{center}
\scalebox{0.8}{\includegraphics{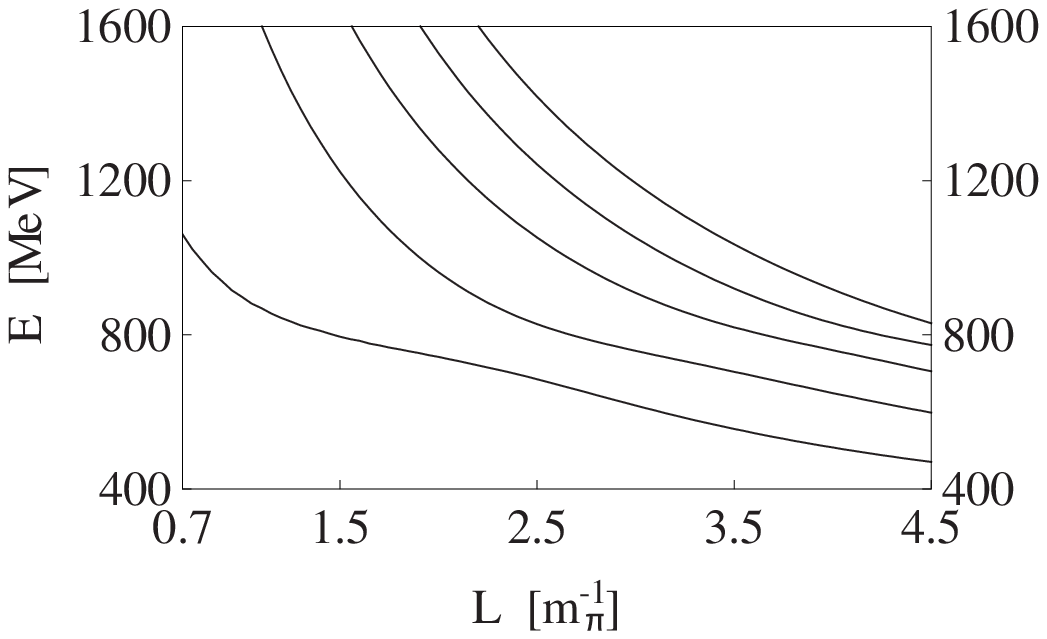}}
\scalebox{0.8}{\includegraphics{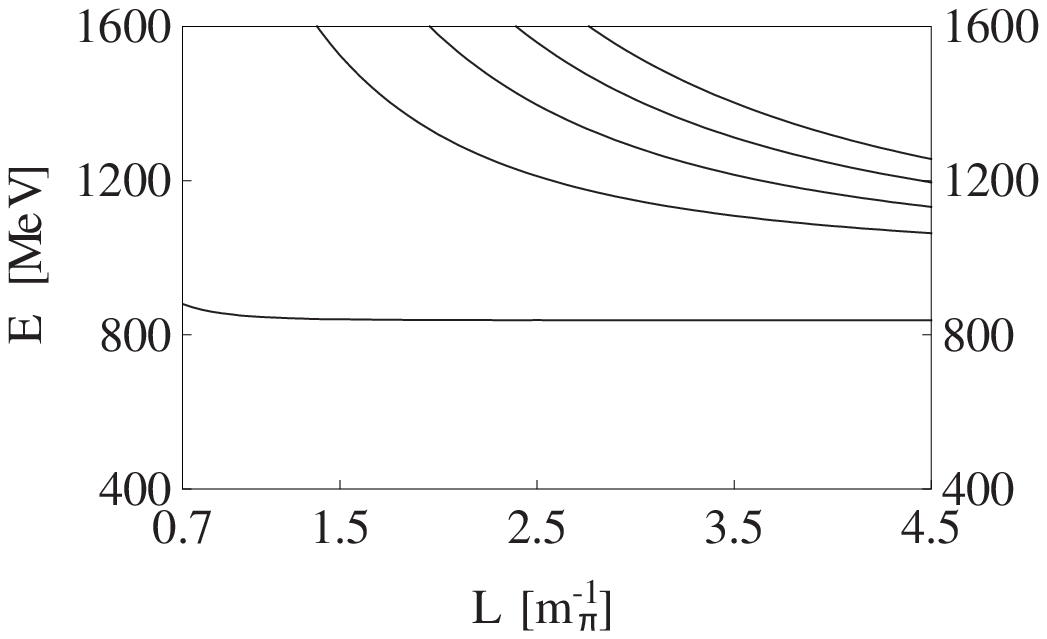}}
\caption{Energy levels as functions of the cubic box size $L$ for the $\pi \pi$ channel and the $K \bar K$ channels, shown in the left and right hand sides, respectively.} \label{fig:ELOne}
\end{center}
\end{figure}

In the left panel of Fig.~\ref{fig:ELMix} we show the energy levels as function of $L$ for the coupled channels, which are obtained after performing an average for different $q_{\rm max}$ values between 1200 MeV and 2000 MeV. On the right panel we show the same ones separately  for values of $q_{\rm max} = 1300$, $1500$, $1700$ and $1900$ MeV. We find that the energy levels for different $q_{\rm max}$ are almost the same. At least we can not differentiate them in Fig.~\ref{fig:ELMix}. This surprising result is simply a consequence of the fact that the $\rho$ meson is very weakly tied to the $\pi \pi$ loops, accounted for by the $\tilde G$ function, where the $L$ dependence appears. We note that the higher curves seem to cross  each other. This does not happen because of the ordering chosen for the levels and then produces the ``avoided level crossing'', but they are a consequence of the coupled channel dynamics. To see it clearly, we  show the results for single $\pi \pi$ and $K \bar K$ channels in Fig.~\ref{fig:ELOne}.

In Fig.~\ref{fig:ELOne} we show the energy levels for the single channel, either the $\pi \pi$ channel or the $K \bar K$ one. We find that the first and second energy levels in the left figure, which are obtained using only  the $\pi \pi$ channel, are quite similar to the results for the coupled channels. However, the higher levels are a mixture of both the $\pi \pi$ and the $K \bar K$ channels. This is quite reasonable since the $K \bar K$ channel does not contribute much in the low energy region. Since the first (lowest) energy level should be well calculated using the chiral unitary approach which is very accurate in this energy region, in the following discussions we shall mainly concentrate on it, and consequently some of our calculations will be done only in the $\pi \pi$ channel.

From these energy levels, we can obtain their corresponding phase shifts. To do this we use only the $\pi \pi$ channel and follow the procedure used in Ref.~\cite{misha}. Since the energy levels are obtained by solving the poles of Eq.~(\ref{eq:EL}), they satisfy the following relation:
\begin{eqnarray}
1 - {\bf V} \tilde {\bf G} \xrightarrow[]{\mbox{one channel}} 
V_{11}(E) \big ( V_{11}(E)^{-1} - \tilde G_{11}(E,L) \big ) = 0 \, ,
\label{eqnlevel}
\end{eqnarray}
where $E= \sqrt s$ is the $\pi \pi$ energy in the CM frame, which can be used to calculate the scattering matrix:
\begin{eqnarray}\label{eq:T11}
T_{11}(E, L) &=& (V_{11}(E)^{-1} - G_{11}(E))^{-1} = ( \tilde G_{11}(E,L) - G_{11}(E))^{-1} \, .
\end{eqnarray}
Since this relation holds for the energy levels shown in Fig.~\ref{fig:ELMix} and Fig.~\ref{fig:ELOne}, $L$ and $E$ are not independent. Moreover when we only consider the first (lowest) energy level, we have $T_{11}(E, L) = T_{11}(E, L(E)) = T_{11}(E)$. From this calculated $T$ matrix one can evaluate the $\pi \pi$ $P$-wave phase shifts which are related to our $T$ by
\begin{eqnarray}\label{eq:delta}
T_{11}(E) &=& { - 8 \pi E \over p \cot\delta(p) - i p } \, ,
\end{eqnarray}
with $p$ the pion CM momentum.

\begin{figure}[hbt]
\begin{center}
\scalebox{0.8}{\includegraphics{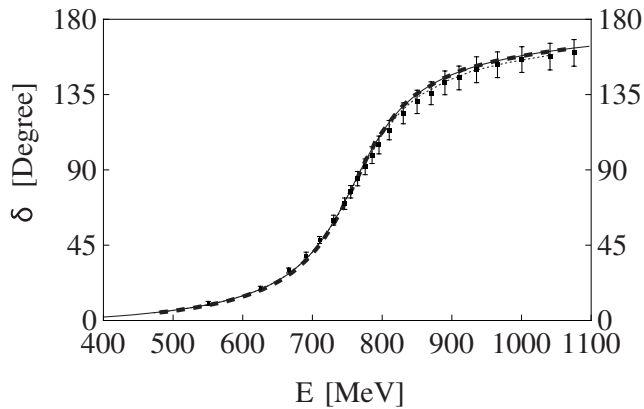}}
\caption{The solid curve is the $\pi \pi$ scattering $P$-wave phase shift calculated using the chiral unitary approach~\cite{formfactor}. The dashed curve is the phase shift extracted from the first energy level of the left figure of Fig.~\ref{fig:ELOne}, obtained using only the $\pi \pi$ channel. The dotted curve is the phase shift extracted from the first energy level of the left figure of Fig.~\ref{fig:ELMix}, obtained using coupled channels. The data are from \cite{Estabrooks:1974vu}.} \label{fig:PhaseRead}
\end{center}
\end{figure}

In Fig.~\ref{fig:PhaseRead}, we show the $\pi \pi$ $P$-wave phase shifts $\delta$, calculated through three different approaches. The solid curve is taken from Ref.~\cite{formfactor} using the chiral unitary approach. This result is consistent with the experimental data~\cite{Estabrooks:1974vu}. The dashed curve is the phase shift extracted from the first energy level of the left figure of Fig.~\ref{fig:ELOne} obtained using only the $\pi \pi$ channel. The results are the same as the solid curve.
The dotted curve is the phase shift extracted from the first energy level of the left figure of Fig.~\ref{fig:ELMix}, obtained with the coupled channels but using only $\pi \pi$ in the analysis. We find that there are small differences in the high energy region, which are caused by the $K \bar K$ channel.

Using the phase shifts $\delta(E)$ we can fit the physical quantities for the $\rho$ meson, such as $m_\rho$, $g_{\rho \pi \pi}$ and $\Gamma_\rho$. We note that $m_\rho$ is the $\rho$ mass we obtained, i.e., one of our outputs; while $M_\rho$ is the bare $\rho$ mass, i.e., one of our inputs. We use the following two equations to extract the $\rho$ properties:
\begin{eqnarray} \label{eq:rhoWidth}
\cot \delta(s) = {m_\rho^2 - s \over \sqrt s \Gamma_\rho(s)} \, ,
~~~{\rm and }~~~
\Gamma_\rho(s) = {p^3 \over s} {g^2_{\rho \pi \pi} \over 8 \pi} \, .
\end{eqnarray}
We note that the factor $8 \pi$ in the second equation is our normalization, while in Ref.~\cite{sasa} the authors use $6 \pi$.

The results from fitting the $\pi \pi$ channel (dashed curve shown in Fig.~\ref{fig:PhaseRead}) are
\begin{eqnarray}
m_{\rho} = 768.6 {\rm~MeV}\, , g_{\rho \pi \pi}=6.59 \, , \Gamma_\rho = 135 {\rm~MeV} \, ,
\label{rhomass1}
\end{eqnarray}
while those from fitting the coupled channels (dotted curve shown in Fig.~\ref{fig:PhaseRead}) are
\begin{eqnarray}
m_{\rho} = 769.6 {\rm~MeV}\, , g_{\rho \pi \pi}=6.79 \, , \Gamma_\rho = 144 {\rm~MeV} \, .
\label{rhomass2}
\end{eqnarray}
This should be compared with the ``exact'' results with the chiral unitary approach with two channels
\begin{eqnarray}
m_{\rho} = 767.9 {\rm~MeV}\, , g_{\rho \pi \pi}=6.69 \, , \Gamma_\rho = 139 {\rm~MeV} \, .
\label{rhomass3}
\end{eqnarray}

\begin{figure}[hbt]
\begin{center}
\scalebox{0.8}{\includegraphics{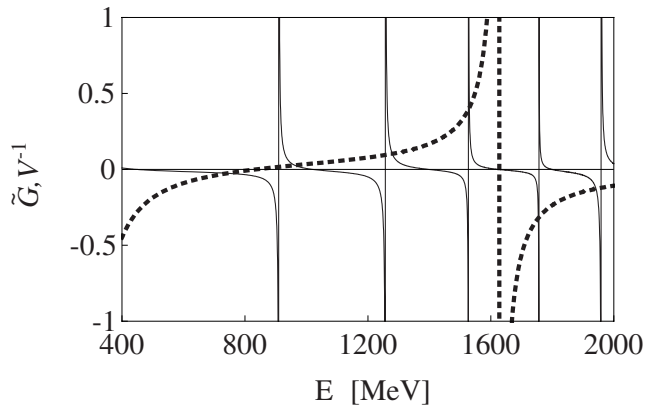}}
\caption{The dashed curve is the $\pi \pi$ inverse potential $ V_{11}(E)^{-1}$ calculated using the chiral unitary approach~\cite{formfactor}. The solid line is $\tilde G$.} \label{fig:cutvgtilde}
\end{center}
\end{figure}

It is interesting to show explicitly how the levels appear in this case. As shown in Eq. (\ref{eqnlevel}), the energies are obtained for one channel by solving the equation $\big ( V_{11}(E)^{-1} - \tilde G_{11}(E,L) \big ) = 0$. In Fig. \ref{fig:cutvgtilde} we show $V_{11}(E)^{-1}$ and $\tilde G_{11}(E,L)$ separately and the points where the two curves cut each other provide the eigenenergies of the box. We observe that $\tilde G$ is a small quantity compared with other hadron scattering cases, since it raises very fast close to the poles which are given by the free energies of the box induced by the boundary condition on the sides of the box. The $\pi \pi $ potential is large, and even has a pole, the CDD pole at the bare mass of the $\rho$. Consequently, $  V_{11}(E)^{-1}$ is also a small quantity, passing through zero around the bare $\rho$ mass. The combination of these two facts leads to an eigenenergy of the box with interaction in the vicinity of the bare $\rho$ mass, which defines the first level. For all the other levels the eigenenergies are very close to the free energies and this tells us that they carry small information on the potential. All these features are a consequence of the peculiar nature of the $\rho$ which is not a composite state of the two pions and cannot be described as a scattering resonance from an ordinary potential, but is a genuine state (basically a $q \bar q$ state) which decays into two pions. Even then the L\"uscher formalism, or our version of it, can be applied to obtain the $\rho$ properties from the $\pi \pi$ phase shifts, as we are doing here.

\section{Dependence on the Pion Mass}
\label{sec:pimass}

In this section, we study the mass and decay width of the $\rho$ meson using different pion masses. There has been pioneering work along these lines using effective theories in \cite{Bruns:2004tj}, evaluating the $\rho$ self-energy with infrared regularization, but the results depend on four low energy constants that ultimately are fitted to lattice data.

We define $m_\pi^0$ to be the physical $\pi$ mass, and now $m_\pi$ is a free parameter. It changes from $m_\pi^0$ to $3 m_\pi^0$ in our study. When changing $m_\pi$, other parameters in our previous formulae, especially the parameter $f$ ($f_\pi$) used in Eqs.~(\ref{eq:Vmatrix}), can also change at the same time. The variation of $f_\pi$ as a function of  $m_\pi$ has been calculated using the IAM in \cite{Pelaez:2010fj,Hanhart:2008mx} and it compares favorably with lattice QCD calculations \cite{Boucaud:2007uk,Beane:2007xs,Noaki:2008gx}. In \cite{nicola,nebreda} the chiral perturbation formula of \cite{gasserannals} is used and the parameters are fitted to old lattice results. The formula contains terms in $m^2_{\pi}$ and terms in $m^2_{\pi}ln(m_{\pi}/\mu)^2$ with $\mu$ a scaling mass. The softer mass dependence of the logarithm term allows one to make an easy fit to the lattice results of \cite{Boucaud:2007uk,Beane:2007xs,Noaki:2008gx} in terms of a quadratic expression in $m_{\pi}$, and we find a good fit to these results by means of
\begin{equation}
\frac{f_{\pi}(m_{\pi})}{f_{\pi}(m^0_{\pi})}= 1+0.035((\frac{m_{\pi}}{m^0_{\pi}})^2-1),
\end{equation}
with $f_{\pi}(m^0_{\pi})=87.4$~MeV as we needed in our fit to the $\rho$ data.

The coupling $G_V$ with the vector meson dominance assumption of \cite{sakurai,Bando:1987br} is related to $f_\pi$ \cite{Ecker:1989yg}, as $G_V = f_\pi / \sqrt 2$. We still take the value $G_V = 53$ MeV of \cite{formfactor} but assume it to be proportional to $f_\pi$ as a function of $m_\pi$. On the other hand, the bare $\rho$ mass, $M_\rho$ in Eqs.~(\ref{eq:Vmatrix}), provides the link of the theory to a genuine component of the $\rho$ meson, not related to the pion cloud, and we assume it to be  $m_{\pi}$ independent.

\begin{figure}[hbt]
\begin{center}
\scalebox{0.46}{\includegraphics{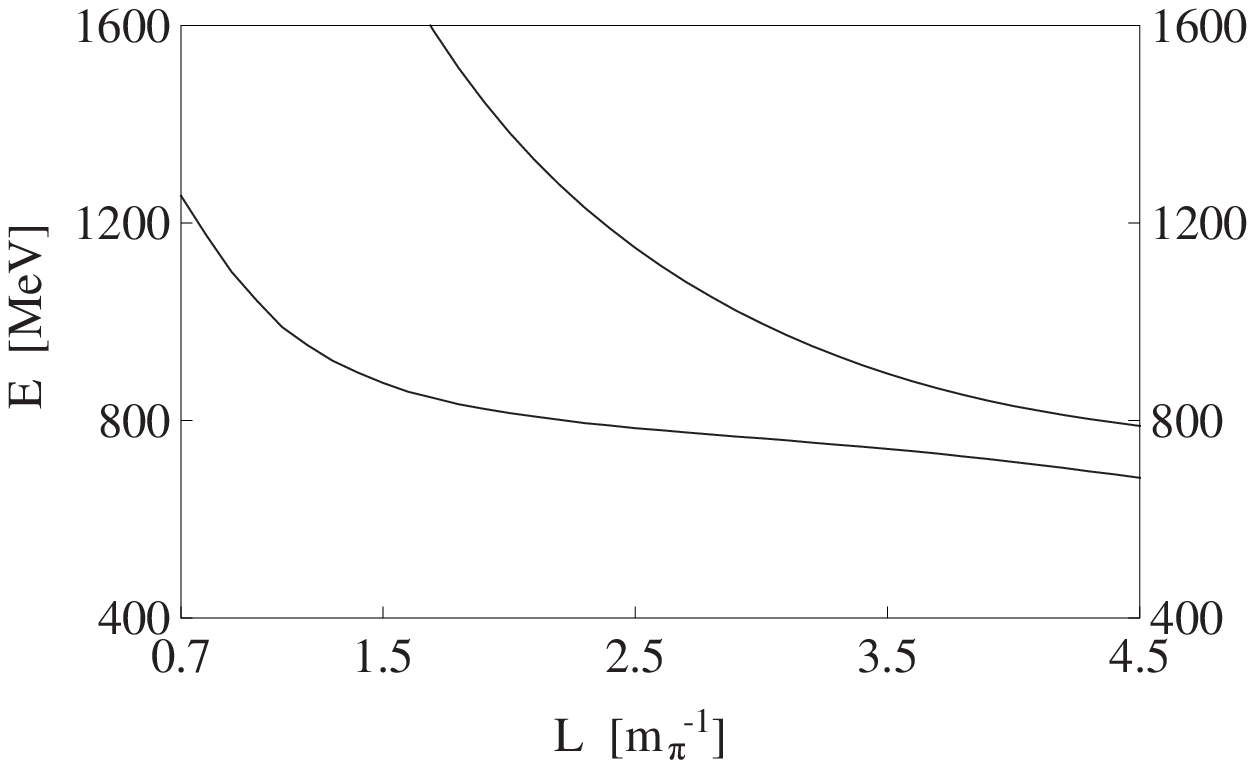}}
\scalebox{0.46}{\includegraphics{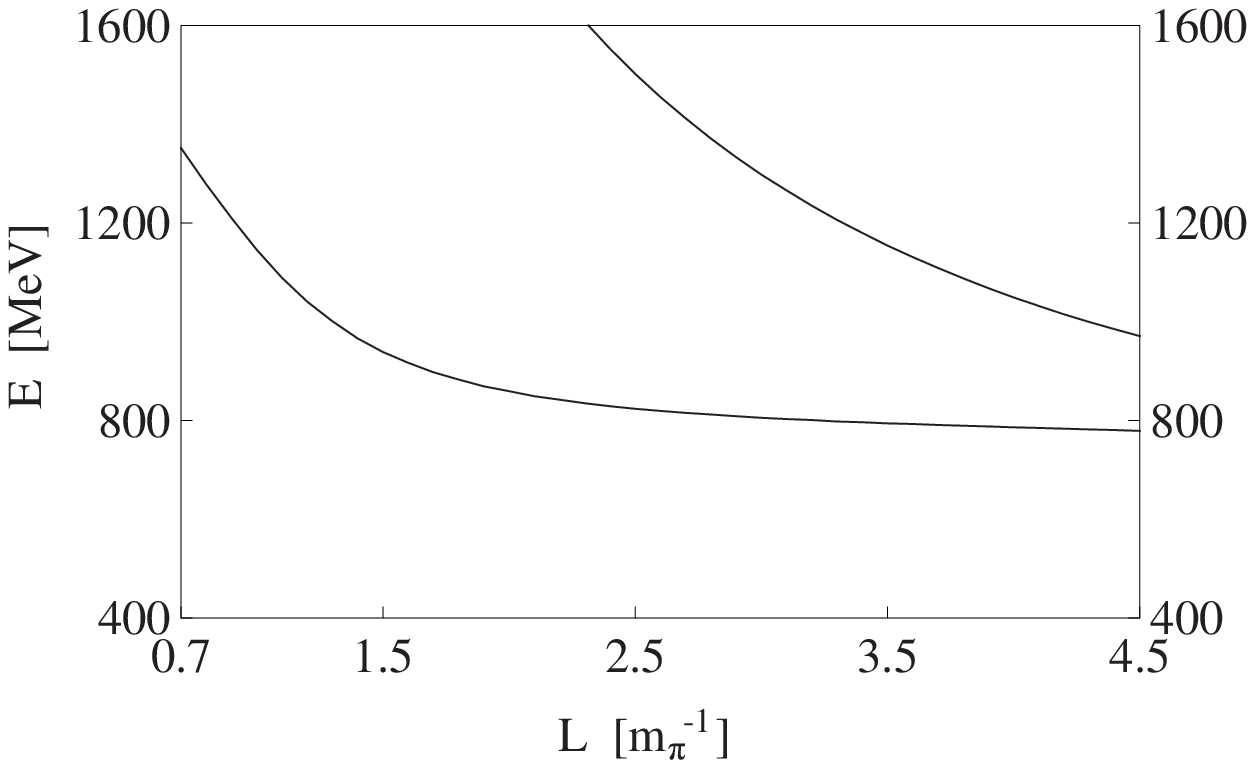}}
\scalebox{0.46}{\includegraphics{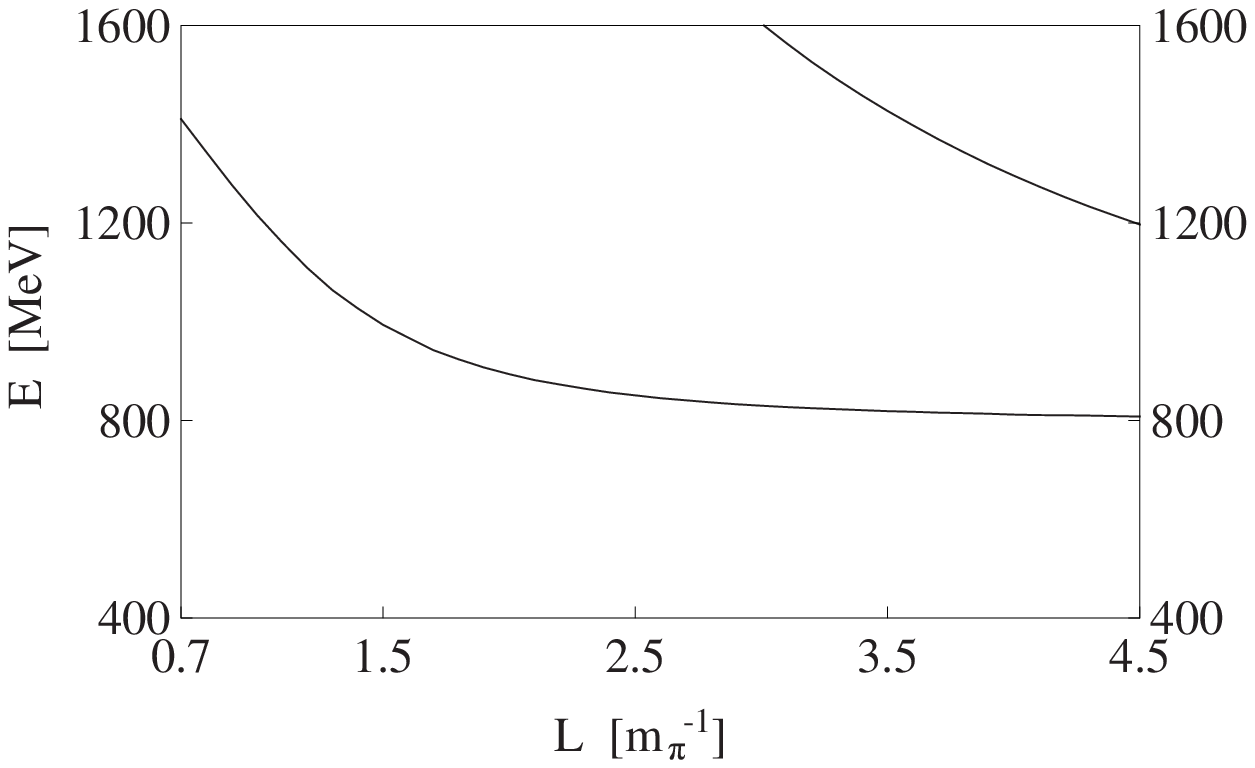}}
\caption{Energy levels as functions of the cubic box size $L$. The left, middle and right figures correspond to $m_\pi = 1.5~m_\pi^0$, $m_\pi = 2.0~m_\pi^0$ and $m_\pi = 2.5~m_\pi^0$, respectively.} \label{fig:mpi}
\end{center}
\end{figure}
Having all the input, we can use the same procedure to calculate the energy levels using different $m_\pi$ values. We show the energy levels using $m_\pi = 1.5~m_\pi^0$ (left), $m_\pi = 2.0~m_\pi^0$ (middle) and $m_\pi = 2.5~m_\pi^0$ (right) in Fig.~\ref{fig:mpi}. We note that the $x$-coordinate is expressed in units of $m_\pi^{-1}$, not $(m_\pi^0)^{-1}$.

We compare our results with the Lattice results of Ref.~\cite{sasa}:
\begin{eqnarray}\nonumber
\begin{array}{c|c}
\hline \hline
\mbox{Our Input and Result} & \mbox{Input and Result in Ref.~\cite{sasa}}
\\ \hline \hline
m_\pi = 276 {\rm MeV}, L = 2.75 m_\pi^{-1} = 1.96 {\rm fm} & m_\pi = 266 {\rm MeV}, L = 1.98 {\rm fm}
\\ \hline
E_1 = 813.2 ^{+7.2}_{-7.3} {\rm MeV}, E_2 = 1390.5 ^{+1.0}_{-1.0} {\rm MeV} & E_1 = 813.4 \pm 6.3 {\rm MeV} , E_2 = 1433.7 \pm 16.1 {\rm MeV}
\\ \hline
\delta_1 = {136.3^\circ} ^{+1.3^\circ}_{-1.4^\circ}, \delta_2 = {175.6^\circ} ^{+0.3^\circ}_{-0.3^\circ} & \delta_1 = 130.56^\circ \pm 1.37 ^\circ , \delta_2 = 146.03^\circ \pm 6.58^\circ
\\ \hline \hline
\end{array}
\end{eqnarray}
We find that the first (lowest) energy level and the extracted phase shift from the two approaches are very similar, while those from the second energy level have small differences. At the end of section~\ref{sec:luscher} a complementary discussion will be made.

\begin{figure}[hbt]
\begin{center}
\scalebox{0.6}{\includegraphics{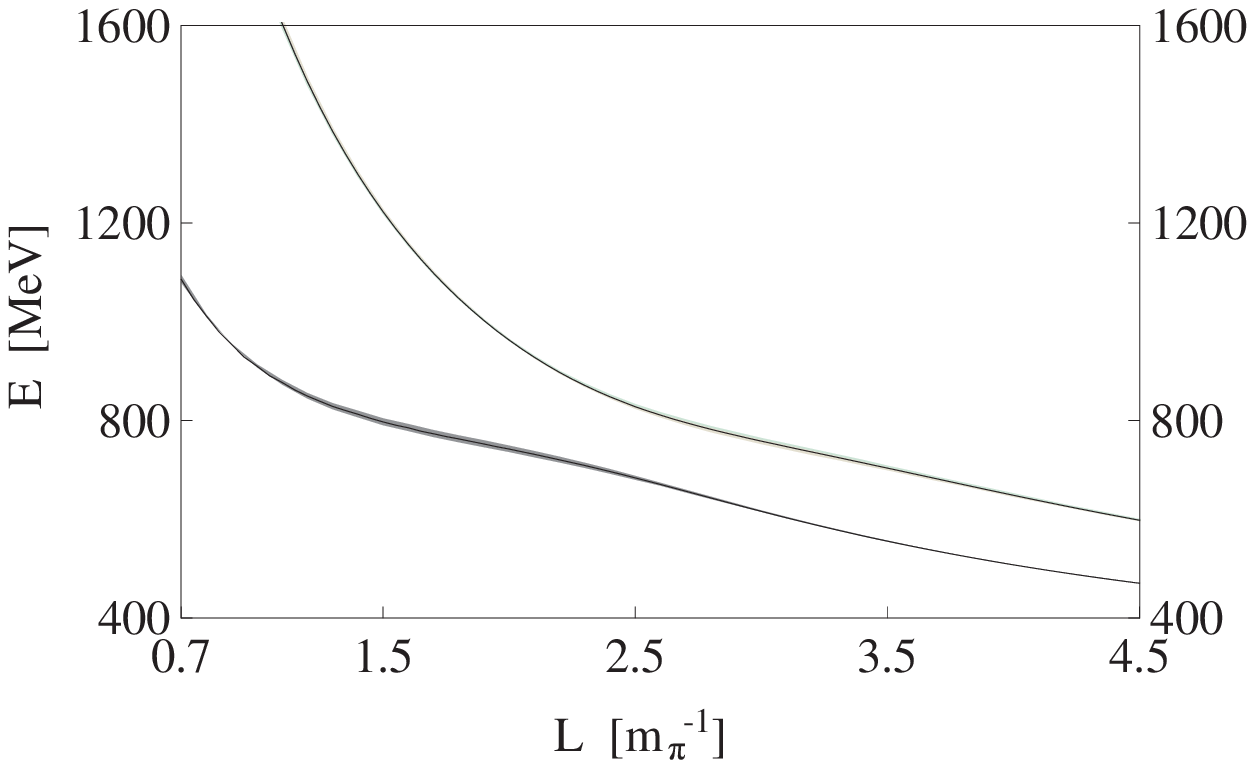}}
\scalebox{0.6}{\includegraphics{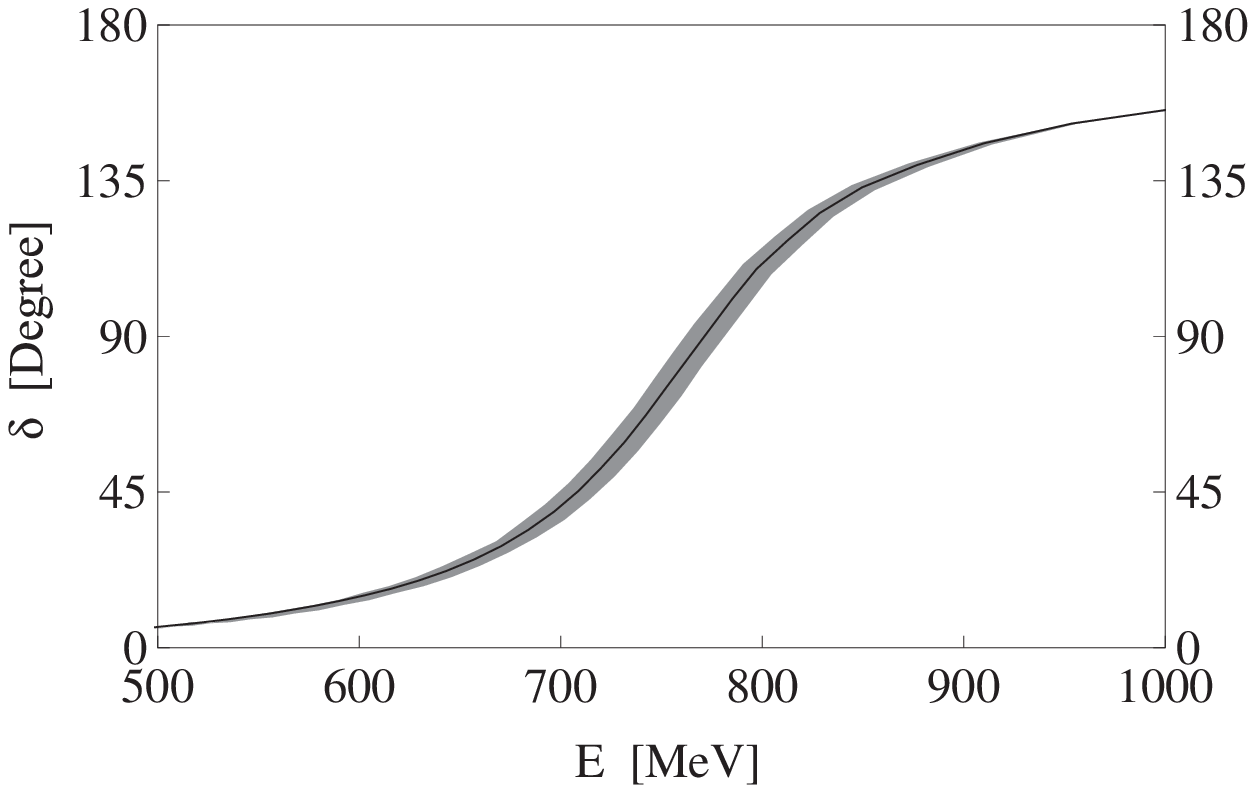}}
\caption{The uncertainty of energy levels and phase shifts.} \label{fig:error}
\end{center}
\end{figure}

In order to estimate the theoretical uncertainties we have proceeded as follows. We change the three parameters $G_V$, $M_\rho$ and $f$ in Eqs.~(\ref{eq:Vmatrix}) and assume that their uncertainty is about 1\%. We can obtain the uncertainty of the energy levels and phase shifts as shown in Fig.~\ref{fig:error}, which is acceptable. Choosing $L = 2.0$ $m_\pi^{-1}$, the uncertainty of energy levels is less than 3\%, and the uncertainty of the induced phase shifts around 700-850 MeV is about 8\%. However, at 815 MeV the errors are smaller and we see from the table above that they are reasonably smaller than the lattice ones. We should keep this in mind for a proper comparison with lattice results.

\begin{figure}[hbt]
\begin{center}
\scalebox{0.8}{\includegraphics{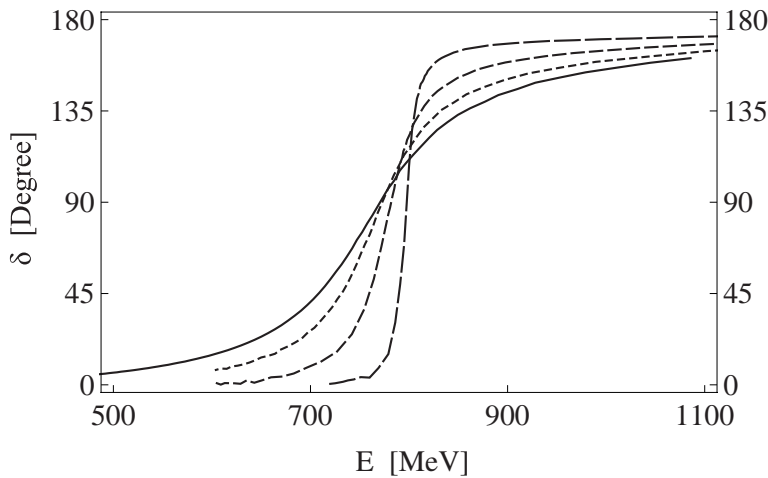}}
\caption{The $\pi \pi$ phase shifts with different pion masses. The solid curve is obtained using physical pion mass, while the short-dashed, middle-dashed, long-dashed curves correspond to $m_\pi = 1.5~m_\pi^0$, $m_\pi = 2.0~m_\pi^0$ and $m_\pi = 2.5~m_\pi^0$, respectively.} \label{fig:ps_mpi}
\end{center}
\end{figure}
Following our previous procedure, we can use these energy levels to obtain the phase shifts. Again we only consider the first (lowest) energy level. The results are shown in Fig.~\ref{fig:ps_mpi}. The solid curve is obtained using the physical pion mass, while the short-dashed, middle-dashed, long-dashed curves correspond to $m_\pi = 1.5~m_\pi^0$, $m_\pi = 2.0~m_\pi^0$ and $m_\pi = 2.5~m_\pi^0$, respectively.

\begin{table}[!hbt]
\caption{The $\rho$ mass, coupling and decay width obtained according to Eq.~(\ref{eq:rhoWidth}).}
\begin{center}
\label{tab:rhomass}
\begin{tabular}{c|cccccccc}
\hline \hline
$m_\pi$ ($m_\pi^0$) & 1.0 & 1.25 & 1.5 & 1.75 & 2.0 & 2.25 & 2.5 & 2.75
\\ \hline
$m_\rho$ (MeV) & $769.6^{+10.2}_{-8.6}$ & 771.0 & 772.8 & 776.6 & $781.8^{+8.4}_{-8.0}$ & 789.1 & 797.5 & 806.5
\\ \hline
$g_{\rho \pi \pi}$ & $6.79^{+0.20}_{-0.25}$ & 6.65 & 6.45 & 6.27 & $6.01^{+0.15}_{-0.15}$ & 5.74 & 5.58 & 5.21
\\ \hline
$\Gamma$ (MeV) & $144^{+9}_{-11}$ & 121 & 96.2 & 72.8 & $49.8^{+3.8}_{-3.4}$ & 30.3 & 15.5 & 4.18
\\ \hline \hline
\end{tabular}
\end{center}
\end{table}
\begin{figure}[hbt]
\begin{center}
\scalebox{0.46}{\includegraphics{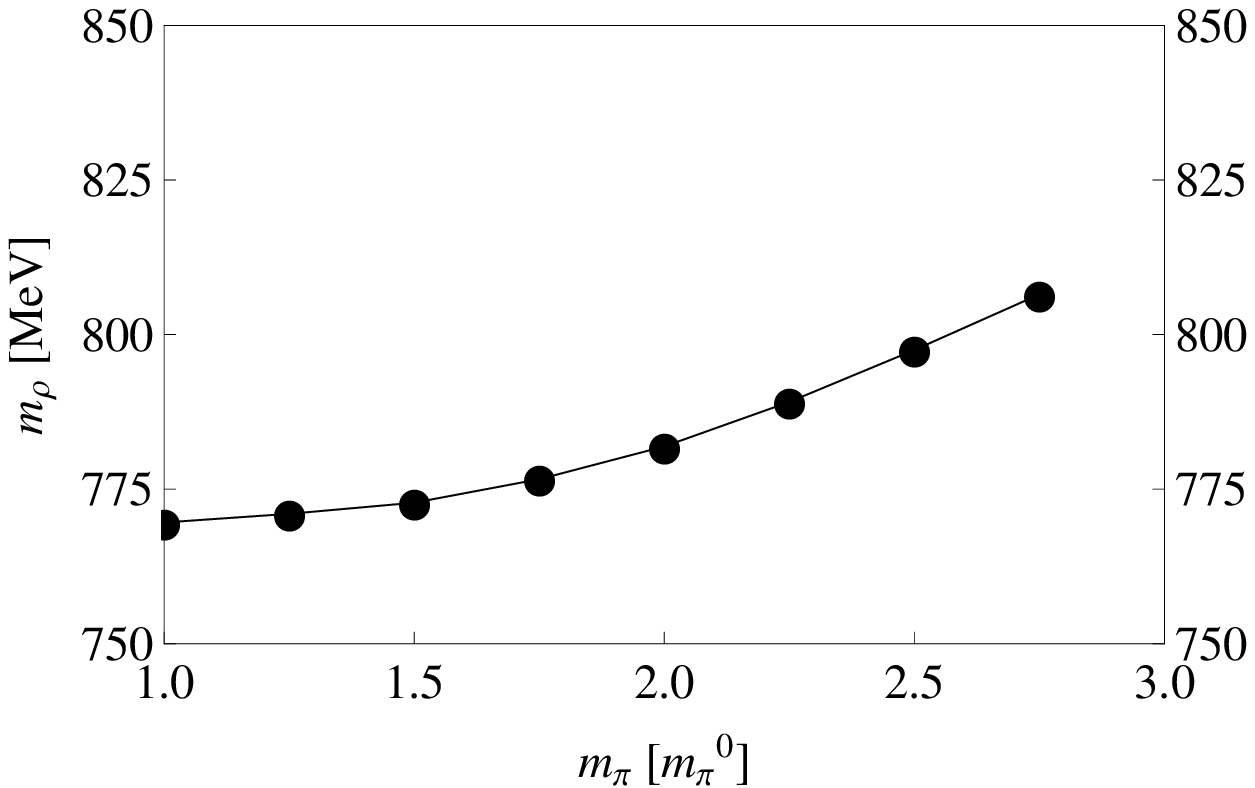}}
\scalebox{0.445}{\includegraphics{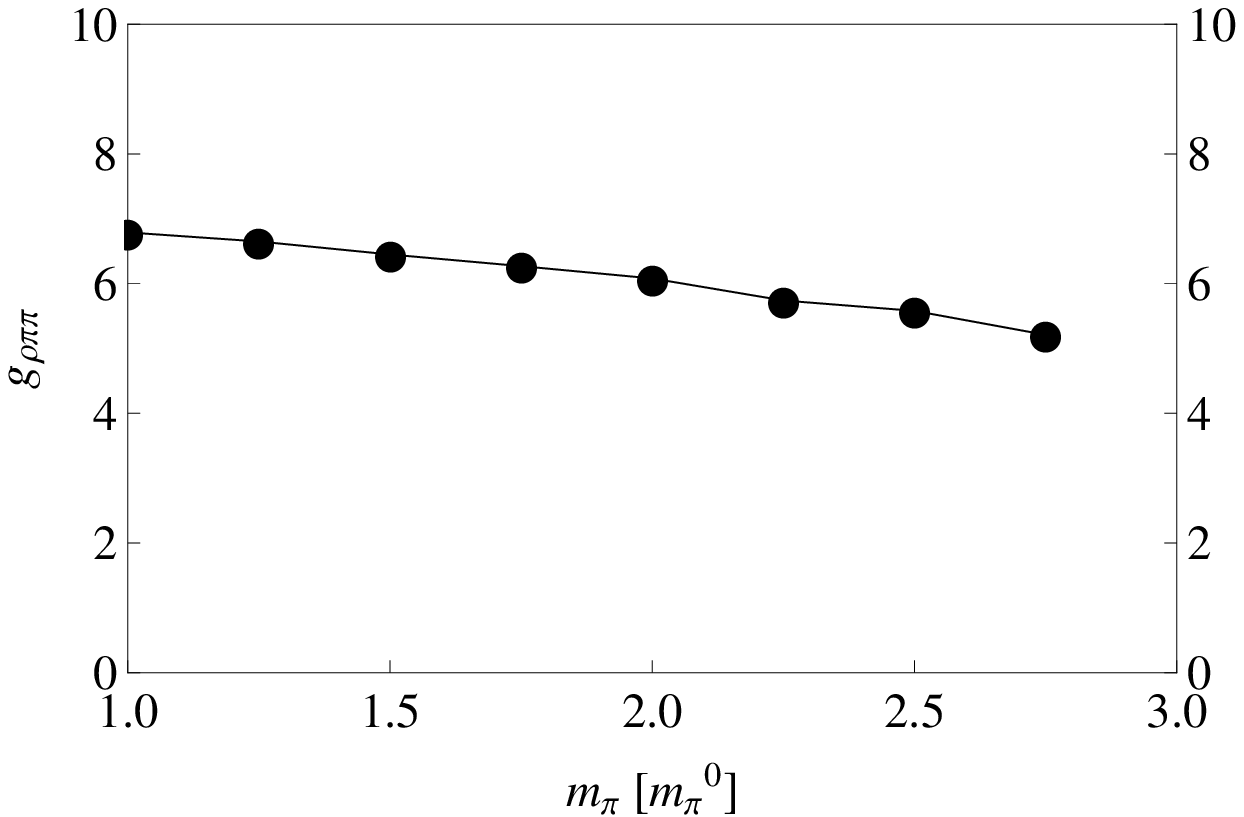}}
\scalebox{0.46}{\includegraphics{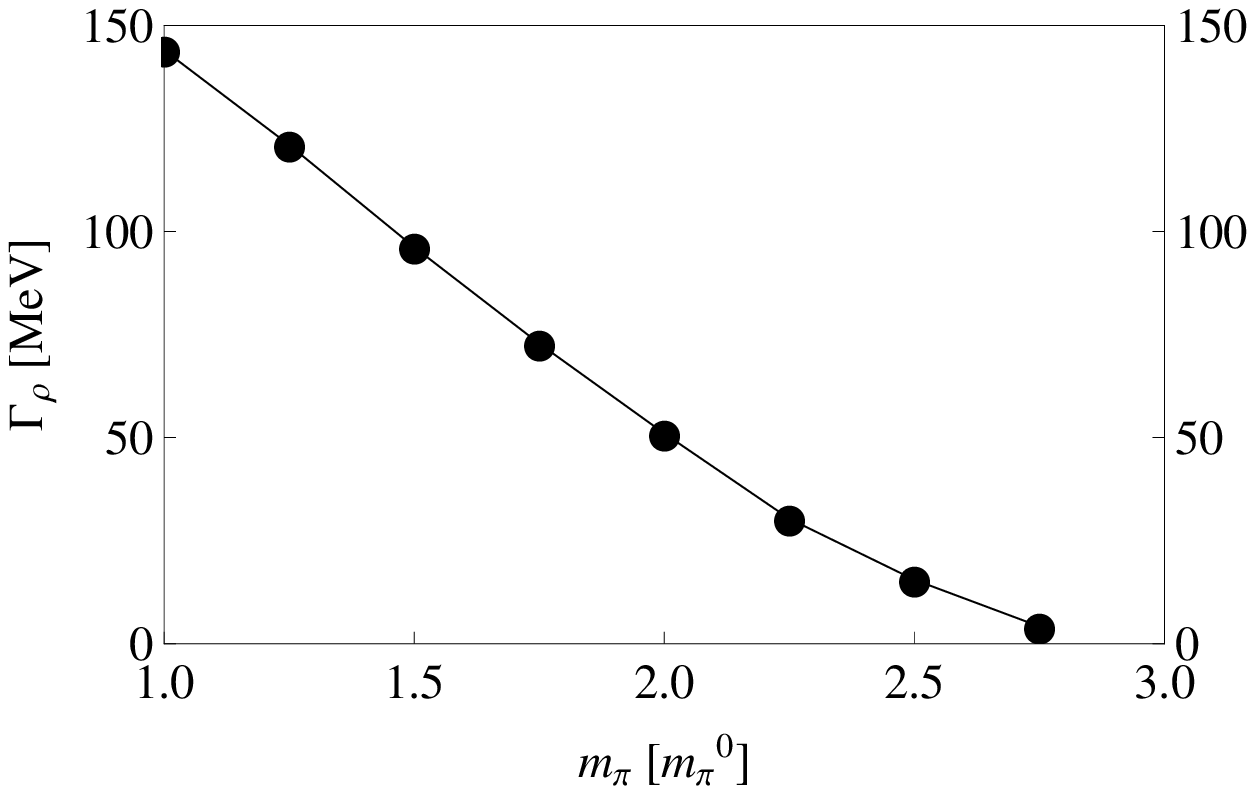}}
\caption{The $\rho$ mass (left), the coupling constant $g_{\rho \pi \pi}$ (middle) and the decay width $\Gamma_\rho$ (right) as functions of $m_\pi$.} \label{fig:rhomass}
\end{center}
\end{figure}

Using the phase shifts shown in Fig.~\ref{fig:ps_mpi}, we can obtain the $\rho$ mass (left), the coupling constant $g_{\rho \pi \pi}$ (middle) and the decay width $\Gamma_\rho$ (right), which are shown in Fig.~\ref{fig:rhomass} as functions of $m_\pi$. We also show the numbers in Table~\ref{tab:rhomass}.
We have also estimated errors for two cases, which give us an idea of these uncertainties in general. They have been evaluated letting the three parameters of the potential, $f, G_V, M_{\rho}$, vary within 1\% such as to still get a very good agreement with the experimental phase shifts. Again we compare our results with the Lattice results in Ref.~\cite{sasa}. They use $m_\pi = 266$ MeV, and the obtained results are $m_\rho = 792\pm7\pm8$ MeV and $g_{\rho \pi \pi} = 5.13\pm0.20$. After correcting $g_{\rho \pi \pi}^2$ for the factor $8\pi \over 6\pi$, the obtained results in \cite{sasa} are $m_\rho = 792\pm7\pm8$ MeV and $g_{\rho \pi \pi} = 5.93\pm0.24$ in our normalization. The obtained $\rho$ mass $792\pm7\pm8$ MeV is in agreement, within errors, with our result $m_\rho(m_\pi = 274 {\rm MeV}) = 781.8^{+8.4}_{-8.0}$ MeV ($m_\pi = 2.0~m_\pi^0$ in Table~\ref{tab:rhomass}), while the obtained coupling constant $g_{\rho \pi \pi} = 5.93\pm0.24$ is also in agreement with our result $g_{\rho \pi \pi}(m_\pi = 274 {\rm MeV}) = 6.01^{+0.15}_{-0.15}$.

\section{Comparison of our results with the standard L\"uscher's Approach}
\label{sec:luscher}

In the Lattice QCD calculations, the L\"uscher approach is used, which is a bit different from ours as discussed in \cite{misha}. To make our analysis complete, especially for the comparison of our results with the Lattice results, in this section we show what one would get from the energy levels obtained here using  the standard L\"uscher's approach. Although the original derivation of L\"uscher formula uses non-relativistic Quantum Mechanics, it is also noted there that its validity can be extended to relativistic field theory. However, we want to show here how in this extension one is making approximations (exponentially suppressed) in the real part of the two particle propagators, which are avoided in our approach. For this we follow the derivation of Ref.~\cite{sachraj} and go to Eqs.~(31) to (36) of this paper. Eq.~(31) of Ref.~\cite{sachraj} is our Eq.~(7) discretized in the box
\begin{eqnarray}
G(p^2) &=& i {1 \over L^3} \sum_{\vec q_i}^{q_{\rm max}} \int {d q^0 \over 2 \pi} {1 \over q^2 - m^2 + i \epsilon} {1 \over (p-q)^2 - M^2 + i \epsilon} \, ,
\end{eqnarray}
where we have substituted  the function $f(q^0,\vec q)$ of ~\cite{sachraj} by a sharp cut-off in three momentum $\theta(q_{\rm max} - |\vec q|)$, which is the one appearing in the chiral unitary approach~\cite{npa}, although the argumentation could be done with a general $f(q^0,\vec q)$ fulfilling the conditions of Ref.~\cite{sachraj}. As in \cite{sachraj} we perform the $q^0$ integration analytically picking up the contribution of the ``particle'' part from the first pole ($q^0 = \omega_q - i \epsilon$) and the ``antiparticle'' part from the second pole ($q^0 = p^0 + \omega_{pq} - i \epsilon$) with $\omega_q = \sqrt{m^2 + \vec q^2}$ and $\omega_{pq} = \sqrt{M^2 + (\vec p - \vec q)^2}$
\begin{eqnarray}
\label{eq:ggg}
G(p^2) &=& {1 \over L^3} \sum_{q_i} \Big \{ {1 \over 2\omega_q}{1 \over p^0 - \omega_q - \omega_{pq} + i \epsilon}{1 \over p^0 - \omega_q + \omega_{pq} + i \epsilon - i \epsilon^\prime}
\\ \nonumber &&+ {1 \over 2\omega_{pq}}{1 \over p^0 + \omega_q + \omega_{pq} }{1 \over p^0 - \omega_q + \omega_{pq} + i \epsilon - i \epsilon^\prime} \Big\} \, ,
\end{eqnarray}
The two terms in the former equation are the two terms that one separates in Eq.~(32) of Ref.~\cite{sachraj}. Note that in these two terms there is a factor, $(p^0 - \omega_q + \omega_{pq} + i \epsilon - i \epsilon^\prime)^{-1}$ which has a pole with unknown position in the complex plane, one does not know where the pole is, in the upper or lower side of the plane. Since the problem is well defined, this indefinition should be fallacious, and indeed, one can see that in the sum of the two terms this denominator cancels exactly the expression of the numerator and the fallacious pole disappears, leading to the expression used in Ref.~\cite{npa}
\begin{eqnarray}
G(p^2) &=& {1 \over L^3} \sum_{q_i} I(q_i) \, ,
\end{eqnarray}
with $I(q_i)$ given by Eq.~(\ref{ifun}) (adding $+i\epsilon$ in the denominator).

For the finite box this unknown sign of the $i \epsilon - i \epsilon^\prime$ does not matter because the pole is never reached in the discrete sum. Hence, the separation of these two terms in Ref.~\cite{sachraj} is justified. However, to reach the formula of L\"uscher an approximation is done in \cite{sachraj} since in the discrete sum of Eq.~(\ref{eq:ggg}) the first term is kept as a discrete sum, but the second term is approximated by an integral (see Eqs.~(35) and (36) of \cite{sachraj}).
The approximation is justified in \cite{sachraj} since poles only appear in the first term of Eq. (\ref{eq:ggg}) and not in the second term, for $\vec p = 0$ and equal masses, hence, substituting the discrete sum for the second term by an integral introduces only exponentially suppressed corrections. Yet, the approach followed here, making a discrete sum of the two terms of Eq. (\ref{eq:ggg}), avoids this unnecessary approximation. Note that problems appear in this separation for $\vec p \neq 0$ or different masses of the particles, since none of the terms is defined, only the sum.

Another way to show the approximations involved in L\"uscher approach is made in Ref.~\cite{misha}. Indeed renaming $\omega_q = \omega_1$, $\omega_{pq} = \omega_2$, the function $I(q)$ of Eq.~(\ref{ifun}) can be written as
\begin{eqnarray}\label{luscherap}
{1 \over 2 \omega_1 \omega_2} {\omega_1 + \omega_2 \over E^2 - (\omega_1 + \omega_2)^2 + i \epsilon} &=& {1 \over 2E} { 1 \over p^2 - \vec q^2 + i \epsilon } - {1 \over 2 \omega_1 \omega_2} {1 \over \omega_1 + \omega_2 + E}
\\ \nonumber && - {1 \over 4 \omega_1 \omega_2} {1 \over \omega_1 - \omega_2 - E} - {1 \over 4 \omega_1 \omega_2} {1 \over \omega_2 - \omega_1 - E} \, ,
\end{eqnarray}
where $p = {{\lambda^{1/2}(E^2, m_\pi^2, m_\pi^2)} / 2 E}$ and $\lambda(x,y,z) = (x-y-z)^2 - 4yz$. In the standard L\"uscher's approach only the first term of the right hand side of this equation is kept. This guarantees that the imaginary part of the propagator is kept exactly, but real terms (which are actually exponentially suppressed in $L$) are neglected. We want to show here the effect of neglecting these terms in the analysis that we have done.

\begin{figure}[hbt]
\begin{center}
\scalebox{0.8}{\includegraphics{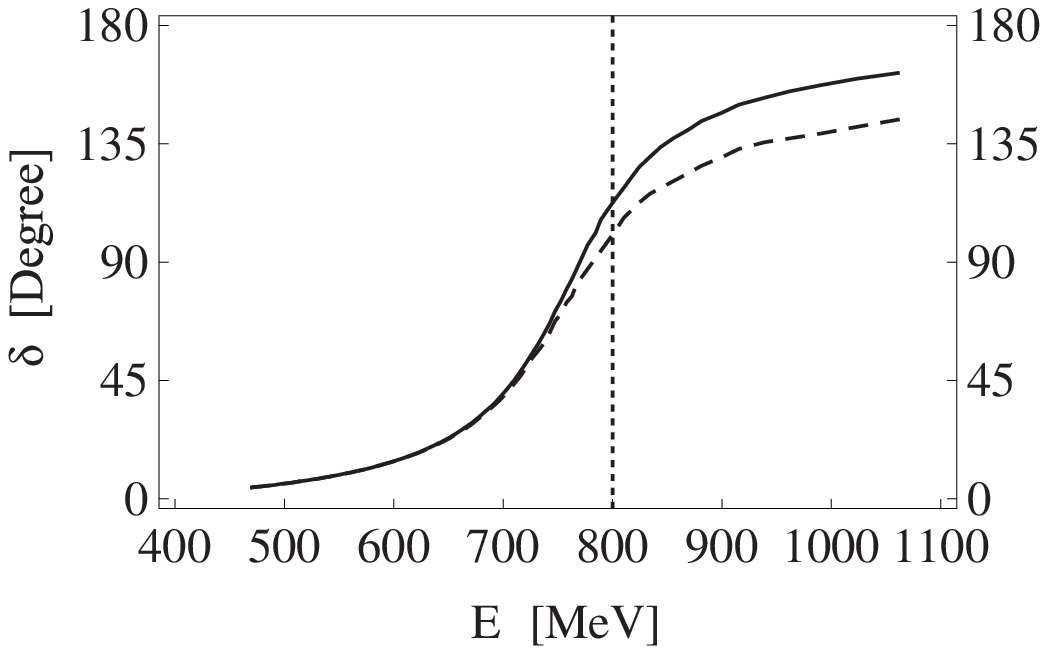}}
\scalebox{0.8}{\includegraphics{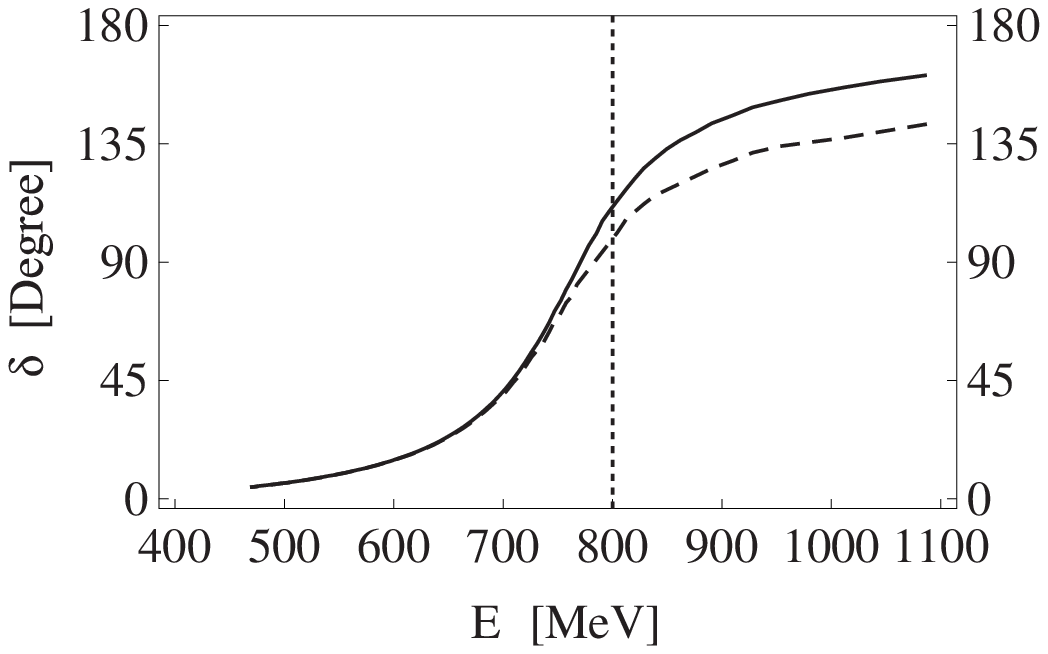}} \caption{The comparison between our result (solid curves) with the standard L\"uscher's approach (dashed curves), respectively. The curves at the left and right hand side are extracted from the first energy level using one-channel and two-channel approach, respectively. The line at 800~ MeV indicates phase shifts obtained with $L m_{\pi}=1.5$ and is the upper energy where both methods should be applied using Eq. (\ref{eq:T11}) and the lowest level.} \label{fig:Luscher}
\end{center}
\end{figure}
We obtain L\"uscher's approach using our formalism, changing $I(q)$ by the first term of  Eq. (\ref{luscherap}). We use as input the energy levels which we have calculated in Sec.~\ref{sec:energylevels} using our $G$-functions (Eqs.~(\ref{eq:GDR}) and (\ref{gtilde})), and then obtain the phase shifts as discussed before but using our expression for $I(q)$ or the approximate one that leads to  L\"uscher's approach. Since we have two sets of energy levels, for the single $\pi \pi$ channel and for the coupled channels, we take these two sets. The results are quite similar for the single and two channel cases and we show the results in Fig.~\ref{fig:Luscher}. The left and right figures are for single channel and coupled channels ``synthetic data'', respectively, extracted with just the $\pi \pi$ channel analysis. We find that in the low energy region the results using the L\"uscher approach or ours are quite similar, while the differences become larger as the energy gets bigger. The new results with the L\"uscher's approach using the single $\pi \pi$ ``synthetic data'' are
\begin{eqnarray}
m_{\rho} = 786.6 {\rm~MeV}\, , g_{\rho \pi \pi}=7.66 \, , \Gamma = 189 {\rm~MeV} \, ,
\label{lushermass1}
\end{eqnarray}
while those obtained from the coupled channels ``synthetic data'' are
\begin{eqnarray}
m_{\rho} = 789.5 {\rm~MeV}\, , g_{\rho \pi \pi}=7.94 \, , \Gamma = 204 {\rm~MeV} \, .
\label{lushermass2}
\end{eqnarray}

By comparing these results with those of Eqs.~(\ref{rhomass1}),~(\ref{rhomass2}) and (\ref{rhomass3}), we can see that there are quite some differences from using one or the other approach. One should note that the phase shifts for energies around 900-1000 MeV are obtained in our case from data with $L$ rather small, of the order of $m_{\pi}^{-1}$ or smaller. In the work of \cite{sasa} small sizes are avoided by getting data using instead the $\pi \pi$ system in moving frames for which the formalism is also available in \cite{Rummukainen:1995vs,sachraj,arXiv:1108.5371,arXiv:1110.0319}, and in
\cite{mishamoving,luisroca} using the formalism of \cite{misha}. This avoids having to go to small values of $L$ to get higher energies. Yet, it would be interesting to see how far down in $L$ one can go and we discuss it below.

As noted before, in order to get the phase shifts of Fig.~\ref{fig:Luscher} at $E \sim 1100$~MeV from the first level, we had to go to values of $L$ close to $0.7~m_{\pi}^{-1}$ in Fig.~\ref{fig:ELMix} left. This is a value too low, where pion size and polarization effects of the pions will certainly play a role, invalidating L\"uscher approach or our related one. It is interesting to quote textually from Ref. \cite{Luscher:1990ux}: {\it It is quite obvious that a discussion of two-particle states and scattering wave
functions in finite volume is only meaningful if polarization effects can be neglected. Essentially what one requires
is that the box is large enough to contain two particles together with their
polarization clouds. In QCD one expects that values of L greater than about 3 fm
($2~m_{\pi}^{-1}$) will do. But there are no general rules as to which is the minimal acceptable box size.}

The formalism that we have used offers an interesting perspective on this observation. Indeed, Eq. (\ref{bethesal}) is derived in Refs. \cite{nsd,ollerulf} using a dispersion relation that requires the on-shell potential in momentum space, but it only includes the right hand cut singularity. It sums meson-meson loops in the $s$-channel.  The polarization comes from loops in the $t$ and $u$ channels, that Eq. (\ref{bethesal}) does not contain and which are volume dependent. Actually, through subtraction constants in the dispersion relation (constant terms in the $G$-function of Eq.~(\ref{Ggasser})), one takes into account an energy average of the left hand cut contribution in
Eq. (\ref{bethesal}), but one misses its volume dependence in the finite box.

As stated above from Ref. \cite{Luscher:1990ux}, there is no general rule for which is the minimal acceptable box size. Fortunately, an answer to this question has become available in the related problem of $\pi \pi $ scattering in s-waves in \cite{rios}, using the IAM \cite{Dobado:1996ps,Oller:1998hw} method or an improved Bethe Salpeter approach \cite{guo}. In \cite{rios} it is shown that one can go down to $L=1.5~m_{\pi}^{-1}$ and the errors induced by neglecting the volume dependence of the left hand cut are still smaller than those studied here by the assumed $10$ MeV errors in the determination of the lattice data that we will consider in the coming section.  The work of \cite{rios} has been extended to the p-wave $\pi \pi$ scattering in the $\rho$ region, the problem studied here, with similar conclusions as for the s-wave \cite{newrios}. This means in practice that we should not go below  $L=1.5~m_{\pi}^{-1}$ in our approach, and this would limit the energies in Fig.~\ref{fig:Luscher} to about 800 MeV. This is already an energy where the differences of the two approaches are already seen, and the slope around the $\rho$ mass changes appreciably, producing a larger width in L\"uscher approach according to Eq.~(\ref{eq:rhoWidth}) (see also Fig. \ref{fig:ps_mpi}, Table \ref{tab:rhomass} and Eqs. (\ref{lushermass1}), (\ref{lushermass2})).

To finalize this section let us go back to the discussion of section~\ref{sec:pimass} and particularly the comparison with the results obtained in \cite{sasa}. We shall do a different exercise than done there. We will take $E_1$ and $E_2$ obtained in \cite{sasa} and we shall evaluate the phase shifts using the L\"uscher formula, used in \cite{sasa}, and our formula, taking the same $m_\pi$ and $L$ values of \cite{sasa}, $m_\pi = 266$ MeV and $L = 1.98$ fm. The phase shifts predicted are given by Eq.~(\ref{eq:delta}), where $T_{11}(E)$ is given by Eq.~(\ref{eq:T11}) in our formalism and by the same formula in the case of L\"uscher, evaluating $\tilde G(s) - G(s)$ by means of Eq.~(\ref{eq:difference}) substituting $I(q)$ by the first term of the right hand side of Eq.~(\ref{luscherap}). We obtain
\begin{eqnarray}
\begin{array}{c|c}
\hline \hline \mbox{Our method} & \mbox{L\"uscher method}
\\ \hline \hline
m_\pi = 266 {\rm MeV}, L = 1.98 {\rm fm} & m_\pi = 266 {\rm MeV}, L = 1.98 {\rm fm}
\\ \hline
E_1 = 813.4\pm6.3 {\rm MeV} , E_2 = 1433.7\pm16.1 {\rm MeV} & E_1 = 813.4\pm6.3 {\rm MeV} , E_2 = 1433.7\pm16.1 {\rm MeV}
\\ \hline
\delta_1 = {133.09^\circ}^{+1.21^\circ}_{-1.21^\circ}, \delta_2 = {154.57^\circ}^{+5.83^\circ}_{-5.91^\circ} & \delta_1 = {130.45^\circ}^{+1.25^\circ}_{-1.26^\circ}, \delta_2 = {154.34^\circ}^{+5.91^\circ}_{-6.06^\circ}
\\ \hline \hline
\end{array}
\end{eqnarray}
The exercise done compares the results with the two methods. The results at the low energy are accurate and similar in both methods. At the higher energy, a more realistic calculation would require in addition to consider the $K \bar K$ channel (with the mass of the $\bar K$ correlated with the one of the pion) and eventually the $4 \pi$ channel (as noted in \cite{sasa}).

It is interesting to remark that $\delta_1$ obtained with the L\"uescher method agrees with the result of ref.~\cite{sasa} quoted in section~\ref{sec:pimass}. For the higher energy $\delta_2$ gives 154$^\circ$ versus 146$^\circ$ in \cite{sasa}. The difference is small but could be surprising in view of the remarkable agreement on $\delta_1$ (130.45$^\circ$ versus 130.56$^\circ$). The answer to this question is found in \cite{sasa} where the authors mention that $m_{\pi}$ is ``roughly'' 266 MeV and $L$ is given ``approximately'' 2.68 $m^{-1}_{\pi}$. We have checked that $\delta_2$ is very sensitive to variations of $m_{\pi}$ and $L$ and we find $\delta_2= 146^\circ$ for $L$=2.69 $m^{-1}_{\pi}$ together with $m_{\pi}=264$ MeV.

\section{The Inverse Process of Getting Phase Shifts from Lattice Data from two levels}

In this section we study the inverse process of getting phase shifts from Lattice Data \cite{misha,alberto,mishakappa,Xie:2012pi,luisroca} using two levels and a parametrized potential. The method was shown to be rather efficient in \cite{misha,alberto,mishakappa,Xie:2012pi,luisroca} (see also \cite{mishakappa} for other possible parameterizations which induce a bigger dispersion on the obtained results). The energy levels in Fig.~\ref{fig:ELMix} and Fig.~\ref{fig:ELOne} obtained in Sec~\ref{sec:energylevels} contain more information than the one that one obtains using only one level, as we have done so far. In this section we assume that they are ``Lattice'' inputs (synthetic data) and analyze them. We shall try to obtain the $V$-matrices as well as the phase shifts from these ``Lattice'' data. The procedure followed here also serves to get an idea of the uncertainties of the phase shifts for a given accuracy of the ``lattice data''.

\begin{figure}[hbt]
\begin{center}
\scalebox{0.83}{\includegraphics{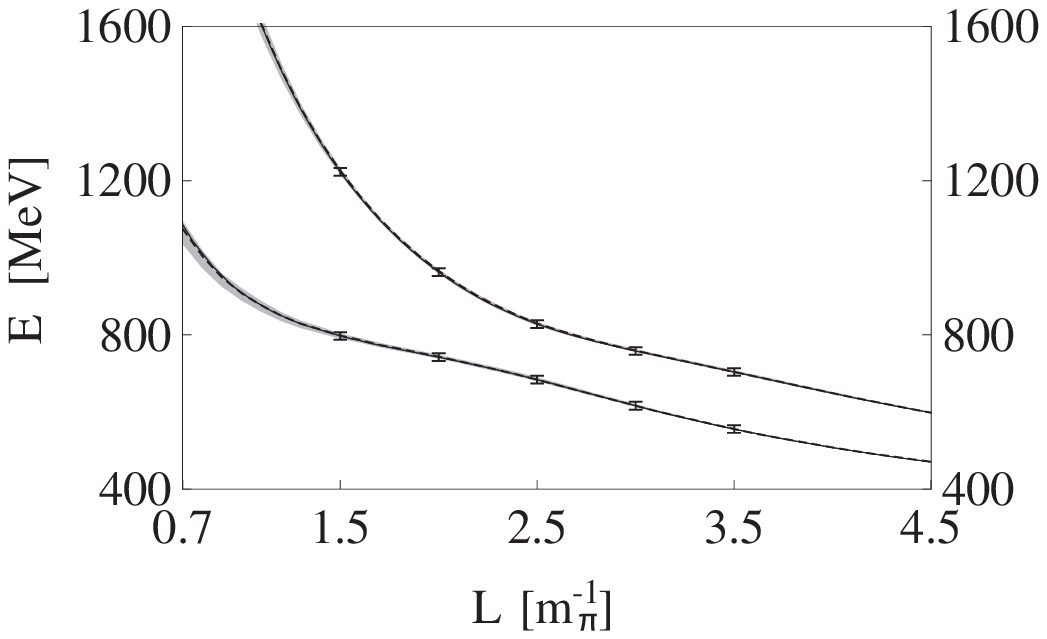}}
\scalebox{0.8}{\includegraphics{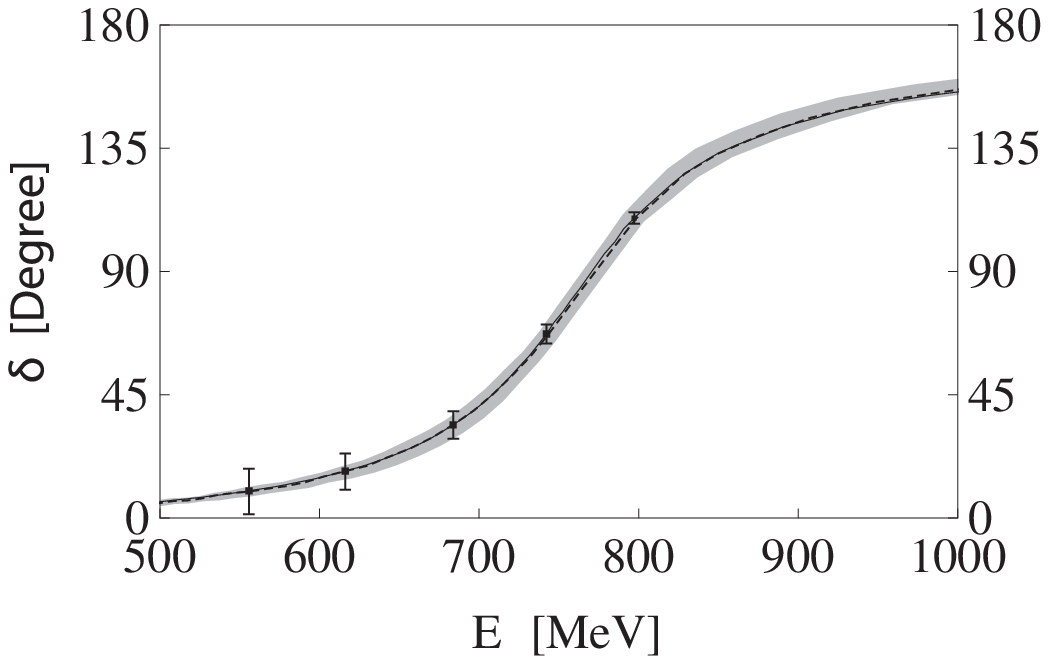}} \caption{Generated data points (10 MeV error, 10 points for each set). The solid curves are the original results, and the dashed curves (shady band) are the fitted results. Fits that fulfill the $\chi^2_{\rm min} + 1$ criterion are also shown in all figures (bands). The $G$-function of Eq.~(\ref{eq:GDR}) is used here. The discrete points in the figure to the right are the results of the direct determination from each ``data'' points on the figure to the left using Eqs.~(\ref{eq:T11}-\ref{eq:delta}) (see text).} \label{fig:ELFitting}
\end{center}
\end{figure}

Since we found that in the region of interest using just the $\pi \pi$ channel provides phase shifts with a high accuracy, we do the one channel analysis and obtain the $V$-matrix. We use the following function which accounts for a CDD pole \cite{Castillejo:1955ed}) to do the one-channel fitting:
\begin{eqnarray}
V_{\pi \pi}(s) = c_1 p_\pi^2 \Big (1 + {c_2 s \over c_3^2 - s} \Big ) \, .
\label{trialpot}
\end{eqnarray}
where $c_1$, $c_2$ and $c_3$ are three free parameters which we shall fit with the ``Lattice'' data shown in Fig.~\ref{fig:ELMix}. We note that these data are calculated using the coupled channels, but, as mentioned, we find that the single $\pi \pi$ channel can already give a good enough fitting. We should mention that with this potential and the formalism of the paper we can accommodate forms of the $T$-matrix which are more general than just a Breit-Wigner. Note that in most lattice calculations the limited amount of phase shifts extracted are extrapolated to other energies assuming a Breit Wigner shape. Our procedure follows Refs.~\cite{misha,alberto}. We take five energies from the first level and five more from the second one, and we associate to them an error of 10 MeV. The following set of parameters leads to a minimum $\chi^2_{\rm min} = 0.15$:
\begin{eqnarray}
c_1 = 9.2 \times 10^{-5} {\rm MeV}^{-2} \, , \, c_2 = 0.73 \, , \, c_3 = 843 {\rm MeV} \, .
\end{eqnarray}
In Fig.~\ref{fig:ELFitting} we show our fitting results calculated from all the possible sets of parameters having $\chi^2 < \chi^2_{\rm min} + 1$. We find that when the energy becomes larger, the error bars become larger at the same time, but the fitting is quite good. Nevertheless, we observe that the percentage in the errors of the induced phase shifts is larger than the one in the value of the energies of the box.

\begin{figure}[hbt]
\begin{center}
\scalebox{0.83}{\includegraphics{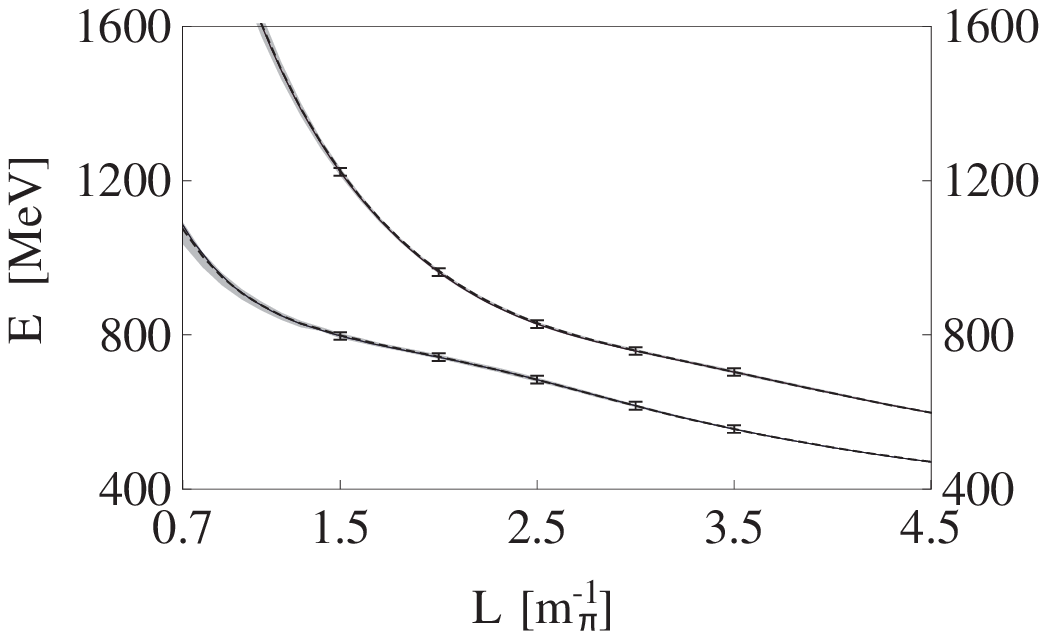}}
\scalebox{0.8}{\includegraphics{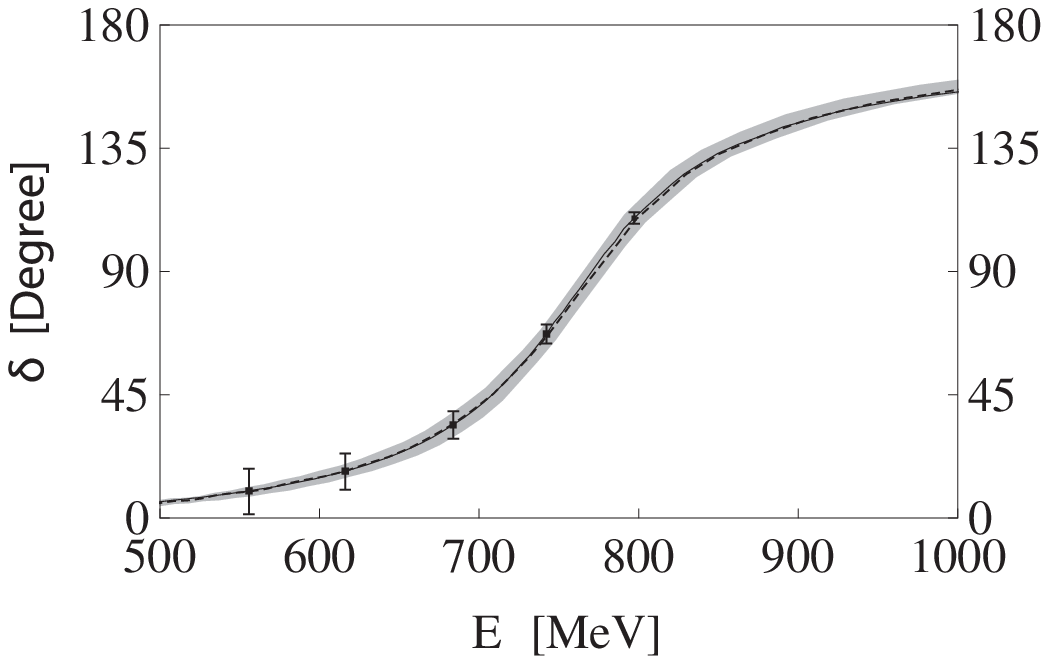}} \caption{Same as in Fig~\ref{fig:ELFitting}. Here we replace the constant -2 of Eq.~(\ref{eq:GDR}) by $-1$.} \label{fig:ELFitCon1}
\end{center}
\end{figure}

\begin{figure}[hbt]
\begin{center}
\scalebox{0.83}{\includegraphics{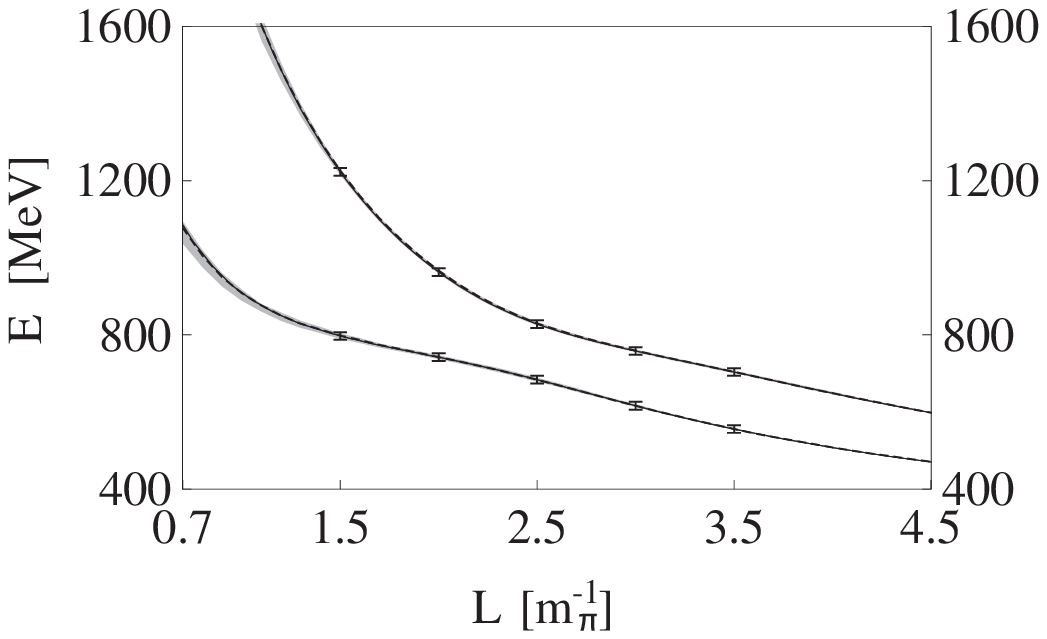}}
\scalebox{0.8}{\includegraphics{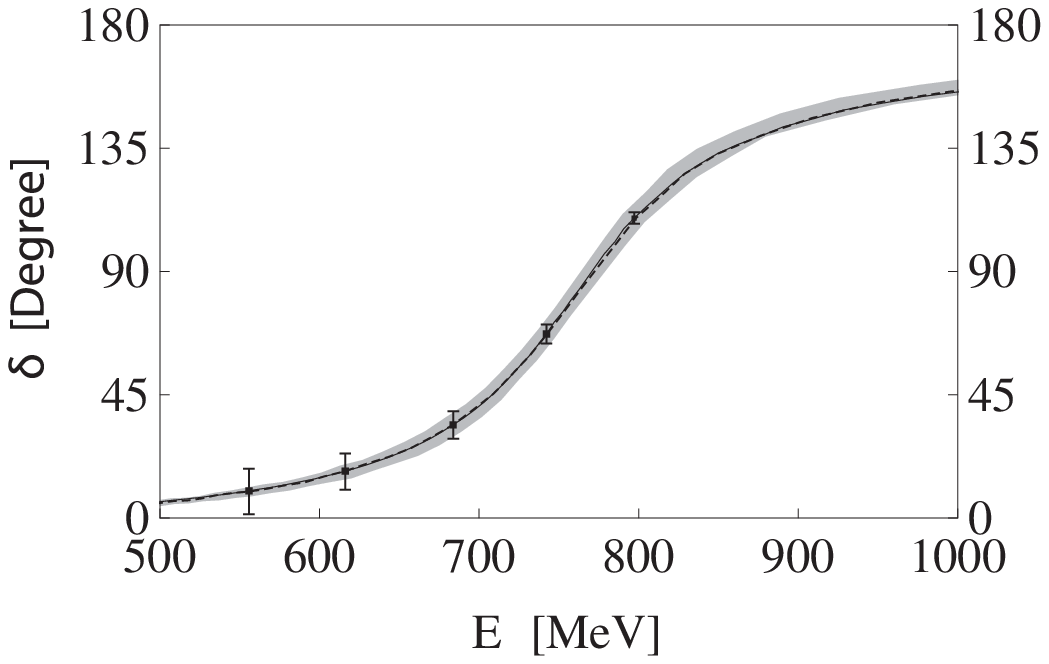}} \caption{Same as in Fig~\ref{fig:ELFitting}. Here we replace the constant -2 of Eq.~(\ref{eq:GDR}) by $-3$.} \label{fig:ELFitCon3}
\end{center}
\end{figure}

In the former analysis we have used the same regularization that was used in the direct problem of getting phase shifts from the chiral unitary approach. However, in a blind analysis one does not know how the $G$-function should be regularized. As mentioned in \cite{misha} the results of this inverse analysis do not depend on which cut off, or subtraction constant one uses in the analysis, as far as one uses the same ones to induce $V$ from the lattice data and then later on to get the phase shifts in the infinite volume from Eq. (\ref{bethesal}). For this purpose we change the constant -2 in Eq. (\ref{eq:GDR}) by -1 and -3 and repeat the procedure. The results are shown in Fig.~\ref{fig:ELFitCon1} and \ref{fig:ELFitCon3}. As one can see, the results for the phase shifts are essentially the same, confirming once more the findings of \cite{misha}.

In Figs.~\ref{fig:ELFitting}, \ref{fig:ELFitCon1} and \ref{fig:ELFitCon3} we see the band of values for the phase shifts and some discrete values with error bars. The band of values has been obtained with a fit to the synthetic data with a potential as explained before ({\it fit method}). The discrete points correspond to the extraction via Eqs. (\ref{eq:T11}-\ref{eq:delta}) of the phase shifts from each individual synthetic data of the five energies of the lower level shown on the left of the figures, with their corresponding error ({\it direct method}). We can see that the errors in the phase shifts are large at small energies but they become smaller as the energy increases. Yet, the overall fit to the whole set of the 10 data from two levels provide a more accurate global determination of the phase shifts than the {\it direct method} in the whole energy range.

\section{Other fit strategies}

In the previous sections we fit the rho mass and width using the energy levels for different volumes in a certain region, such as $0.7~m_\pi^{-1} < L < 4.5~m_\pi^{-1}$ for the case of $m_\pi = m_\pi^0$. We can also use the energy levels for a single volume or two different volumes. In this section we show these analyses, which can serve to plan calculational strategies in actual QCD Lattice simulations. Some examples are:

\begin{enumerate}

\item Taking the lowest 4 energy levels at $L = 3.0~m_\pi^{-1}$ from Fig.~\ref{fig:ELMix}, we can use both the one-channel approach (see the discussions in Sec.~\ref{sec:energylevels}) as well as the standard L\"uscher's approach (see the discussions in Sec.~\ref{sec:luscher}), and the results are
    \begin{eqnarray}
    \nonumber &&\mbox{One-Channel results: } m_\rho = 767.8^{+8.8}_{-9.0}~{\rm MeV}, g_{\rho \pi \pi} = 6.34^{+0.26}_{-0.27}, \Gamma = 125^{+11}_{-11}~{\rm MeV} \, ,
    \\ \nonumber &&\mbox{L\"uscher's approach: } m_\rho = 768.7^{+8.8}_{-9.0}~{\rm MeV}, g_{\rho \pi \pi} = 6.34^{+0.26}_{-0.27}, \Gamma = 125^{+11}_{-11}~{\rm MeV} \, .
    \end{eqnarray}

\item Taking the lowest 4 energy levels at $L = 3.0~m_\pi^{-1}$ and other lowest 4 energy levels at $L = 4.0~m_\pi^{-1}$ from Fig.~\ref{fig:ELMix}, the results are
    \begin{eqnarray}
    \nonumber &&\mbox{One-Channel results: } m_\rho = 768.6^{+9.1}_{-9.3}~{\rm MeV}, g_{\rho \pi \pi} = 6.52^{+0.24}_{-0.23}, \Gamma = 132^{+11}_{-10}~{\rm MeV} \, ,
    \\ \nonumber &&\mbox{L\"uscher's approach: } m_\rho = 769.3^{+9.1}_{-9.3}~{\rm MeV}, g_{\rho \pi \pi} = 6.52^{+0.24}_{-0.24}, \Gamma = 132^{+11}_{-10}~{\rm MeV} \, .
    \end{eqnarray}

As one can see by comparing these results with the ``exact'' results with the chiral unitary approach with two channels, Eq. (\ref{rhomass3}), both the coupling and the width have improved by adding the new four levels.

\item Taking the lowest 4 energy levels at $L = 2.0~m_\pi^{-1}$ from Fig.~\ref{fig:ELMix}, the results are
    \begin{eqnarray}
    \nonumber &&\mbox{One-Channel results: } m_\rho = 768.0^{+8.8}_{-8.8}~{\rm MeV}, g_{\rho \pi \pi} = 6.53^{+0.27}_{-0.26}, \Gamma = 132^{+12}_{-11}~{\rm MeV} \, ,
    \\ \nonumber &&\mbox{L\"uscher's approach: } m_\rho = 772.6^{+8.5}_{-8.6}~{\rm MeV}, g_{\rho \pi \pi} = 6.54^{+0.26}_{-0.27}, \Gamma = 134^{+12}_{-12}~{\rm MeV} \, .
    \end{eqnarray}

  Note that we have obtained results of the same quality as in the former case but using only four levels rather than eight, and with a smaller box size, with the resulting economy of time in actual QCD calculations.

\item Taking the lowest 2 energy levels at $L = 1.5~m_\pi^{-1}$ and other lowest 2 energy levels at $L = 2.0~m_\pi^{-1}$ from Fig.~\ref{fig:ELMix}, the results are
    \begin{eqnarray}
    \nonumber &&\mbox{One-Channel results: } m_\rho = 769.6^{+9.3}_{-9.5}~{\rm MeV}, g_{\rho \pi \pi} = 6.35^{+0.25}_{-0.23}, \Gamma = 126^{+10}_{-10}~{\rm MeV} \, ,
    \\ \nonumber &&\mbox{L\"uscher's approach: } m_\rho = 772.3^{+9.2}_{-9.2}~{\rm MeV}, g_{\rho \pi \pi} = 6.26^{+0.25}_{-0.24}, \Gamma = 123^{+10}_{-10}~{\rm MeV} \, .
    \end{eqnarray}

\end{enumerate}
These results are all consistent with each others as well as our previous results, but as the volume becomes smaller, the differences between our results and the L\"uscher's results become (slightly) larger. So we meet the same situation which we have found in Sec.~\ref{sec:luscher}.

In principle, adding new levels should result into a higher accuracy in the results, but one must be careful since a situation can be reached where the two channel analysis becomes mandatory. Let us see an example of this: in Fig.~\ref{fig:PhaseRead} we show the results using one $\pi\pi$ channel and two coupled channels of $\pi\pi$ and $K \bar K$, and the results are almost the same, as shown in Eqs.~(\ref{rhomass1}) and (\ref{rhomass2}). Therefore, the contribution of the $K \bar K$ channel can be neglected when we only use the lowest energy levels. However, comparing Figs.~\ref{fig:ELMix} and \ref{fig:ELOne}, we can find that several higher energy levels are affected by the $K \bar K$ channel. Here we use an example to show this. We take the lowest 5 energy levels at $L = 3.0~m_\pi^{-1}$ from Fig.~\ref{fig:ELMix}. First we fit the $V$-matrices, only $V^{I=1}_{\pi\pi,\pi\pi}$ for one-channel fitting and $V^{I=1}_{\pi\pi,\pi\pi}$, $V^{I=1}_{\pi\pi,K \bar K}$ and $V^{I=1}_{K \bar K,K \bar K}$ for two-channel fitting. Then we calculate the rho mass and width. The results are:
\begin{eqnarray}
\nonumber &&\mbox{One-Channel Fitting: } \chi^2 = 17, m_\rho = 826\pm61~{\rm MeV}, g_{\rho \pi \pi} = 5.8\pm1.1, \Gamma = 115\pm35~{\rm MeV} \, ,
\\ \nonumber &&\mbox{Two-Channel Fitting: } \chi^2 = 0.005, m_\rho = 767.1^{+9.5}_{-9.1}~{\rm MeV}, g_{\rho \pi \pi} = 6.39^{+0.24}_{-0.26}, \Gamma = 127^{+10}_{-11}~{\rm MeV} \, ,
\end{eqnarray}
where the best fitting has a $\chi^2$ obtained by associating to the lowest 5 energy levels an error of 10 MeV. We see that these two fittings are significantly different, which means that the $K \bar K$ channel can not be neglected in this case. The two-channel fitting is much better, and consistent with our previous results.

\section{Conclusion}
\label{sec:summary}

Our aim has been to provide an efficient strategy to obtain $\pi \pi $ phase shifts, and thus the $\rho$-meson properties from energy levels obtained in lattice calculations. With this purpose in mind,
we have made a study of the $\pi \pi$ interaction in p-wave in a finite box using the chiral unitary approach which is very successful to provide phase shifts and the pion form factor. With this procedure we obtain discrete energies in the box as a function of $L$ and of $m_{\pi}$ which compare favorably with present lattice results. The second part consist of getting the $\pi \pi$ phase shifts in infinite volume from these finite box energies for what we use our equivalent procedure to L\"uscher's approach developed in \cite{misha}, which improves it for small lattice sizes. We use two methods, the first one ({\it direct method}), valid for only one channel, produces the phase shifts without having to use an intermediate potential. The second one uses a parametrized potential and makes a fit to data. This second procedure takes advantage of the correlation among data and provides a global fit that allows one to get phase shifts for all energies and not only for the energies which are eigenvalues of the box, as in the first method, or in L\"uscher's approach. The strategy of the fit method is one of the interesting findings of the work, which produces phase shifts in a large range of energies with good accuracy. Yet, the fact that the $\rho$ is not a composite state of two pions, but has mostly a $q \bar q$ component, forces one to use an unusual "potential", incorporating a CDD pole, Eq. (\ref{trialpot}).

We also estimated errors induced in the phase shifts from the errors in the energies of the box. We could see that by taking a set of about 10 energies from two levels one could induce good phase shifts with reasonable errors, although we noted that the percentage of the errors in the phase shifts was larger than the one in the energies. Typically, errors of 10 MeV in the eigenenergies of the box resulted in large errors for small energies using the {\it direct method} but smaller errors at energies around 800 MeV. The {\rm fit method} provides a better global agreement for all energies and the errors in the phase shifts, induced by the 10 MeV uncertainties in the eigenenergies of the box, are of the order of 6\%.

So far we have used the centroid of the synthetic data centered at the exact curve. In actual lattice data there will also be a dispersion of the centroids. In \cite{misha} this was taken into account and a good statistical study was done. The conclusion, which we can accept also in the present case, is that taking into account a dispersion of the centroids of about 5 MeV, produces extra uncertainties in the results of maximum 50\% of those produced by the 10 MeV uncertainty of the data. Hence we end with a round 10\% errors in our case.

We also showed that if one uses periodic boundary conditions in the rest frame of the  $\pi \pi$ system one has to use small values of the box size in the first level where the standard L\"uscher approach produces too large errors. While this problem has been avoided in some lattice works by obtaining data in a $\pi \pi$ moving frame, we show that using our procedure one can go to smaller sizes with a considerable saving of the computer time in actual QCD lattice calculations. Certainly, one cannot go to very small values of $L$ where polarization effects not considered would invalidate both approaches. However, making use of recent results in $\pi \pi $ scattering where these effects have been evaluated, we can state that the guess of $L>2~m_{\pi}^{-1}$ of \cite{Luscher:1990ux} is indeed safe, and that even going down to $L=1.5~m_{\pi}^{-1}$ induces errors smaller than those found here from the assumed 10 MeV errors in the determination of the lattice energies in the box.
These findings can serve as a guideline for future studies of the $\rho$ meson in these calculations, which would also allow to use pion masses closer to the physical one.

As a byproduct, we have also shown that an analysis of the $E<1000~MeV$ region can be well conducted with only the $\pi \pi $ channel, and the effect of the $K \bar K $ channel is still small. From the perspective of lattice calculations we have also shown for which levels the $K \bar K $ channel will be important and were the "avoided level crossing" will appear, which is due to the interplay of the $\pi \pi $ and $K \bar K $ channels.

We also investigated other fit strategies which can serve as guidelines for future QCD lattice calculations. We found that using four levels for a single volume can serve to make an accurate determination of the $\rho$ properties, even using only the $\pi \pi $ channel. Yet, increasing the number of levels does not necessarily improve the results, since one may bump in a region where the $K \bar K $ channel is important and then the use of the coupled channels in the analysis becomes mandatory. In this case we also showed that the analysis of the lattice results with the two channel formalism, which is equally simple, provides again accurate results on the $\rho$ properties.

\section*{Acknowledgments}

We would like to thank B.~X.~Sun for help in the calculations at the early stage of the problem. Useful discussions with M.~Savage are much appreciated. A careful reading and comments from U.-G.~Mei\ss ner, A.~Martinez Torres, M.~D\"oring and L.~Roca are much appreciated. This work is partly supported by DGICYT contract FIS2011-28853-C02-01, the Generalitat Valenciana in the program Prometeo, 2009/090, and the EU Integrated Infrastructure Initiative Hadron Physics 3 Project under Grant Agreement no. 283286, the National Natural Science Foundation of China under Grant No. 11147140 and No. 11205011, and the Scientific Research Foundation for the Returned Overseas Chinese Scholars, State Education Ministry.

\end{document}